\documentclass[12pt]{article}

\usepackage[left=2.8cm,right=2.8cm,top=2.8cm,bottom=2.8cm]{geometry}

\usepackage[colorlinks=true,linkcolor=black,citecolor=teal,urlcolor=MidnightBlue,filecolor=black]{hyperref}

\usepackage{amsmath}
\usepackage{amsfonts}
\usepackage{amssymb}
\usepackage{stmaryrd}
\usepackage{mathtools}
\usepackage{mathrsfs}
\usepackage{scalerel}
\usepackage[dvipsnames]{xcolor}
\definecolor{SchoolColor}{rgb}{0.6471, 0.1098, 0.1882} % Crimson
\usepackage[nosort]{cite}
\usepackage[normalem]{ulem}
\usepackage{multirow,booktabs}
\usepackage[colorinlistoftodos,prependcaption,textsize=tiny]{todonotes}
\usepackage{filecontents}
\usepackage{ wasysym } % For the sun symbol
\usepackage{eucal} % Nicer \mathcal{}

\providecommand{\Lt}{{\tt L}}
\renewcommand{\Lt}{{\tt L}}
\providecommand{\Mt}{{\tt M}}
\renewcommand{\Mt}{{\tt M}}
\providecommand{\Jt}{{\tt J}}
\renewcommand{\Jt}{{\tt J}}
\providecommand{\Pt}{{\tt P}}
\renewcommand{\Pt}{{\tt P}}
\providecommand{\Gt}{{\tt G}}
\renewcommand{\Gt}{{\tt G}}
\providecommand{\Rt}{{\tt R}}
\renewcommand{\Rt}{{\tt R}}

 \csname
@addtoreset\endcsname{equation}{section}

\definecolor{color1}{rgb}{0.22,0.45,0.70}

\DeclareMathOperator{\extdm}{d}
\newcommand{\extd}{\extdm \!}

\newcommand{\unity}{1\hspace{-0.243em}\text{l}}

\begin{document}

%\begin{flushright}
%\begin{tabular}{c}
%Preprint number
%\end{tabular}
%\end{flushright}

\vspace{30pt}

\begin{center}

%%%%%%%%%%%%%%%%%%%%%%%%%%%%%%%%%%%%%%%%%%%%%%%%%%%%%%%%%%%%%%%%%%%%

{\Large\sc Spectral Flow in 3D Flat Spacetimes}\\

---------------------------------------------------------------------------------------------------------

%%%%%%%%%%%%%%%%%%%%%%%%%%%%%%%%%%%%%%%%%%%%%%%%%%%%%%%%%%%%%%%%%%%%

\vspace{25pt}
{\sc R.~Basu${}^{\; a,\,b}$, S.~Detournay${}^{\; a}$ and M.~Riegler${}^{\; a}$}
\vspace{10pt}

${}^a$\sl\small
Physique Th{\'e}orique et Math{\'e}matique\\
Universit{\'e} libre de Bruxelles and International Solvay Institutes\\
Campus Plaine C.P. 231\\
B-1050 Bruxelles, Belgium
\vspace{10pt}

\vspace{10pt}
{${}^b$\sl\small
Saha Institute of Nuclear Physics\\
Block - AF, Sector - 1\\
Bidhan nagar, Kolkata 700064,\ India\\
\vspace{2pt}
and \\
\vspace{2pt}
%${}^c$\sl\small
Theoretische Natuurkunde, \\
Vrije Universiteit Brussel, \\
Pleinlaan 2, B-1050 Brussels, Belgium

{\it rudranil.basu@ulb.ac.be,\\ sdetourn@ulb.ac.be, \\ max.riegler@ulb.ac.be} 
}

%%%%%%%%%%%%%%%%%%%%%%%%%%%%%%%%%%%%%%%%%%%%%%%
\vspace{50pt} {\sc\large Abstract} \end{center}

\noindent
In this paper we investigate spectral flow symmetry in asymptotically flat spacetimes both from a gravity as well as a putative dual quantum field theory perspective. On the gravity side we consider models in Einstein gravity and supergravity as well as their ``reloaded'' versions, present suitable boundary conditions, determine the respective asymptotic symmetry algebras and the thermal entropy of cosmological solutions in each of these models. On the quantum field theory side we identify the spectral flow symmetry as automorphisms of the underlying symmetry algebra of the theory. Using spectral flow invariance we then determine the thermal entropy of these quantum field theories and find perfect agreement with the results from the gravity side. In addition we determine logarithmic corrections to the thermal entropy.

%%%%%%%%%%%%%%%%%%%%%%%%%%%%%%%%%%%%%%%%%%%%%%%

\newpage

%%%%%%%%%%%%%%%%%%%%%%%%%%%%%%%%%%%%%%%%%%%%%%%%%%%%%%%%%%%%%%%%%%%%

\tableofcontents
\hypersetup{linkcolor=SchoolColor}
\newpage

%--------------------------------------------------
\section{Introduction}
%--------------------------------------------------

The holographic principle has been one of most groundbreaking ideas in theoretical physics during the last two decades, its most successful realization being in the form of the Anti-de Sitter (AdS)/conformal field theory (CFT) correspondence \cite{Maldacena:1997re}. Given the huge success of the AdS/CFT correspondence, it is natural to explore holography in other contexts beyond AdS/CFT in order to broaden the scope of the principle. In this paper we want to focus in particular on the concept of flat space holography. This aims at establishing the holographic principle for gravitational dynamics in asymptotically flat spacetimes. A possible theory of quantum gravity in such a holographic scenario is expected to have a dual field theoretic description on the co-dimension one, null hypersurface at asymptotic infinity.

In the context of flat space holography, it is often useful to keep in mind the AdS/CFT correspondence in lower dimensions. To be more precise, one should refer to the case of three-dimensional asymptotically AdS spacetimes. In the seminal work (much before the advent of AdS/CFT correspondence in the string theoretic framework) of Brown and Henneaux \cite{Brown:1986nw} the asymptotic symmetry algebra of three-dimensional AdS spacetimes was shown to coincide with the symmetries of a two-dimensional conformal field theory. Working along similar lines, the asymptotic symmetry algebra for three-dimensional asymptotically flat spacetimes was worked out in \cite{Barnich:2006av}\footnote{ This work extended previous considerations of \cite{Ashtekar:1996cd}.} and shown to be the three-dimensional Bondi-Metzner-Sachs algebra ($\mathfrak{bms}_3$) \cite{Bondi:1962px,Sachs:1962zza}. The Lie algebra corresponding to $\mathfrak{bms}_3$ can also be understood as an algebraic contraction of the conformal algebra in two dimensions \cite{Bagchi:2010eg}. In a physical sense this contraction corresponds to an ultrarelativistic limit of the conformal symmetry generators \cite{Bagchi:2012cy}. The resulting Lie algebra consists of a semidirect product of a single copy of Virasoro algebra with an Abelian ideal. Much of the progress in flat holography until now makes use of this symmetry algebra as the basic symmetries of a putative dual field theory, identifying physically interesting structures in the bulk with physical quantities on the quantum field theory side such as e.g. \cite{Fareghbal:2013ifa,Bagchi:2014iea,Barnich:2015mui,Bagchi:2015wna,Basu:2015evh,Campoleoni:2015qrh}. Explaining properties such as the thermal entropy of flat space cosmologies\footnote{These cosmological solutions in asymptotically flat spacetimes were first described in \cite{Cornalba:2002fi,Cornalba:2003kd}.} (FSC) \cite{Barnich:2012xq, Bagchi:2012xr} is one such success.

The pristine example of three-dimensional AdS/CFT mentioned above considers the gravitational dynamics to be governed by pure Einstein gravity. Various ramifications of it introduced more complex dynamics both in terms of the geometric content such as e.g. adding a gravitational Chern--Simons (CS) term \cite{Grumiller:2008es, Henneaux:2009pw, Skenderis:2009nt, Afshar:2011qw} or other higher derivative interactions \cite{Hohm:2010jc, Sinha:2010ai}, inclusion of matter interactions \cite{Henneaux:2002wm,Fujisawa:2013ima,Arvanitakis:2014yja} as well as varied boundary conditions \cite{Compere:2013bya,Troessaert:2013fma,Avery:2013dja,Troessaert:2015gra,Donnay:2015abr,Afshar:2015wjm,Afshar:2016wfy,Perez:2016vqo,Grumiller:2016pqb}. These modifications alter the asymptotic symmetries and hence the kinematical structure of the putative dual two-dimensional field theory as well. However, for (asymptotically) flat space holography, which is a relatively new area of research, explorations to similar avenues are few\footnote{See e.g. \cite{Afshar:2015wjm,Afshar:2016kjj,Grumiller:2017sjh} for some ramifications in asymptotically flat spacetimes.}.

In a very recent work \cite{Detournay:2016sfv} a new set of boundary conditions for gravitational fields for asymptotically flat spacetimes was introduced such that the asymptotic symmetry algebra of $\mathfrak{bms}_3$ is enhanced by a $\hat{\mathfrak{u}}(1) \oplus \hat{\mathfrak{u}}(1)$ current algebra. When these kinematical symmetries are realized dynamically (governed by pure Einstein gravity or Topologically Massive Gravity (TMG) \cite{Deser:1981wh,Deser:1982vy}), the current algebra receives a  non-trivial level, given in terms of Newton's constant. The present study arose from the curiosity as to how this current can be understood in the putative dual field theory at the boundary and/or if there may be other descriptions of it in the gravity side. 

There are at least two viewpoints to answer this question. One way of thinking is that this extra symmetry in the dual two-dimensional field theory is the R-symmetry of extended supersymmetry (SUSY) on top of $\mathfrak{bms}_3$. In order to follow up this line of thought we concentrate in this work on a particular avatar of $\mathcal{N}=4$ SUSY that contains the Abelian R-symmetry mentioned above. The bulk description then can be determined to be a supergravity theory with appropriate amount of SUSY. It is worthwhile to mention that SUSY extensions of $\mathfrak{bms}_3$ algebra have been explored recently in the literature in various contexts. These include $\mathcal{N}=1$ or $2$ SUSY extensions \cite{Barnich:2014cwa, Barnich:2015sca, Lodato:2016alv, Banerjee:2016nio} of $\mathfrak{bms}_3$ in a flat holography set up. On the other hand the algebra describing residual worldsheet symmetry of tensionless superstrings \cite{Bagchi:2016yyf} or ambi-twistor strings \cite{Casali:2016atr} has been shown to be equivalent to an algebra isomorphic to $\mathcal{N}=1$ super-$\mathfrak{bms}_3$. The other viewpoint calls for a $\hat{\mathfrak{u}}(1) \oplus \hat{\mathfrak{u}(1)}$ gauge field in the bulk sourced by this new current at the boundary. This second picture requires introduction of a Chern-Simons term in the bulk action which on the other hand, does not bring along any new degrees of freedom. We have successfully implemented these two realizations in this article.

While trying to understand the role of this new bosonic current from the field theory side, we also recall an interesting aspect of certain superconformal field theories. Superconformal symmetry with extended SUSY in two dimensions, expressed via the super Virasoro algebra, possesses an inner automorphism called ``spectral flow''. While this is a statement of invariance at the level of the algebra, the representation theory is affected as well in the sense that highest weights defining a module shift linearly under a spectral flow. On the other hand, even in absence of SUSY, this invariance is seen in certain Virasoro-Kac Moody systems as well \cite{Maldacena:2000hw}. Apart from just being an algebraic curiosity, this property has also been successfully exploited in the case of AdS$_3$ holography. The examples range from calculating entropy of charged black holes \cite{Kraus:2006wn, Detournay:2012pc} to analyzing elliptic genera corresponding to CFTs having well posed gravity duals in AdS$_3$ \cite{Benjamin:2015hsa}.

One of the key observations of the present work is that we have found spectral flow invariance for $\mathfrak{bms}_3$ algebras augmented with $\hat{\mathfrak{u}}(1) \oplus \hat{\mathfrak{u}}(1)$ currents, both with and without SUSY. In the context of discussing spectral flow for the $\mathfrak{bms}_3$ algebra with the above mentioned current structure, it is natural to ask whether such an automorphism is observed even if one includes only a single $\hat{\mathfrak{u}}(1)$ current. The answer again is in the affirmative, at least when we do not include superymmetry. 

Moreover we use this automorphism for Einstein gravity, supergravity in flat space as well ``reloaded'' versions thereof to calculate the thermal entropy, including the logarithmic corrections, for flat space cosmologies. Apart from the usual attributes of mass and angular momentum, the horizon now carries charges corresponding to SUSY R-charge or the charges due to the extra gauge field in the two pictures mentioned above. As a result the expressions for the entropy at different orders in quantum corrections are spectral flow invariant.

We organize the paper in the following way. In Section~\ref{sec:GravityToyModel} we present a gravity toy model with spectral flow symmetry in the bulk first in AdS$_3$ and then show how this model can be used to determine similar models in flat space. In Section~\ref{sec:SpectralFlow} we introduce the $\mathfrak{bms}_3$ algebra supplemented with two $\hat{\mathfrak{u}}(1)$ current algebras, both with and without SUSY. We show the spectral flow invariance for the two cases separately. In the next Section~\ref{sec:GravityModels} we present suitable boundary conditions for a number of different gravity models in flat space namely Einstein gravity and supergravity as well as ``reloaded'' versions thereof. Thermodynamic properties of flat space cosmologies from a gravity perspective are presented in Section~\ref{sec:ThermalGravity} while the field theory calculation for the entropy is presented in Section~\ref{sec:ThermalField}. We then conclude by summarizing the key results and presenting some open problems for further studies.

%--------------------------------------------------
\section{A Gravity Toy Model}\label{sec:GravityToyModel}
%--------------------------------------------------

In this section, we illustrate with a very simple toy model how spectral flow symmetry manifests itself in the bulk, first in AdS$_3$ and then show how one can take an appropriate limit to three-dimensional asymptotically flat gravity.\\
Since the simplest example of having spectral flow symmetry in the context of AdS$_3$/CFT$_2$ holography is provided in case the CFT is deformed by at least one $\hat{\mathfrak{u}}(1)$ current algebra a natural model to consider in this context is Einstein gravity with negative cosmological constant in 3D that is coupled to at least one $\mathfrak{u}(1)$ gauge field \cite{Kraus:2006wn}. With a bit of hindsight in regards to a suitable model in asymptotically flat spacetimes we will couple two $\mathfrak{u}(1)$ gauge fields to Einstein gravity in such a way that the asymptotic symmetry algebra is given by two copies of $\mathfrak{vir}\inplus\hat{\mathfrak{u}}(1)$. This can be done in a particular straightforward way in a first order formalism using the Chern-Simons formulation of gravity in 3D \cite{Witten:1988hc}. Thus we consider the difference of two Chern-Simons actions each supplemented with a $\mathfrak{u}(1)$ gauge field i.e.
    \begin{equation}
        I = I_{\textrm{CS}+\mathfrak{u}(1)}[k,A;\kappa,C]-I_{\textrm{CS}+\mathfrak{u}(1)}[k,\bar{A};\bar{\kappa},\bar{C}],
    \end{equation}
with
\begin{equation}\label{eq:ToyCSAction}
		I_{\textrm{CS}+\mathfrak{u}(1)}[k,A;\kappa,C] = \frac{k}{4\pi} \int_{\mathcal{M}} \langle A \wedge \extd A +\frac23 A \wedge A \wedge A\rangle\,+\frac{\kappa}{8\pi}\int_{\mathcal{M}}\left\langle C\wedge\extd C\right\rangle,
	\end{equation}    
where $k=\frac{\ell}{4G_N}$. Here $\ell$ is the AdS radius and $G_N$ Newton's constant in three dimensions. The gauge fields $A$ and $\bar{A}$ take values in $\mathfrak{sl}(2,\mathbb{R})$ whereas the gauge fields $C$ and $\bar{C}$ take values in $\mathfrak{u}(1)$. The $\mathfrak{sl}(2,\mathbb{R})$ generators are labelled by $\mathfrak{L}_n$ with $n=0,\pm1$ and the $\mathfrak{u}(1)$ generator by $\mathfrak{S}$. We chose a basis such that the non-vanishing commutation relations of these generators are given by
    \begin{equation}
        [\mathfrak{L}_n,\mathfrak{L}_m]=(n-m)\mathfrak{L}_{n+m}.
    \end{equation}
The invariant bilinear form denoted by $\langle\ldots\rangle$ in \eqref{eq:ToyCSAction} in this basis is given by
	\begin{subequations}\label{eq:ToySLInvBilForm}
	\begin{align}
		\langle \mathfrak{L}_n\mathfrak{L}_m\rangle & =\left(
			\begin{array}{c|ccc}
				  &\mathfrak{L}_1&\mathfrak{L}_0&\mathfrak{L}_{-1}\\
				\hline
				\mathfrak{L}_1&0&0&-1\\
				\mathfrak{L}_0&0&\frac{1}{2}&0\\
				\mathfrak{L}_{-1}&-1&0&0
			\end{array}\right),\\
		\langle \mathfrak{S}\mathfrak{S}\rangle & =1.	
	\end{align}		
	\end{subequations}
The topology of the manifold is that of a solid cylinder. In addition we choose coordinates such that there is a radial direction $0\leq r<\infty$ and the boundary of the cylinder is parametrized by a time coordinate\footnote{It should be noted that this time coordinate is dimensionless i.e. it is actually the ratio of a dimensionful time coordinate that we again with some hindsight call $u=\ell t$.} $-\infty<t<\infty$ as well as an angular coordinate $\varphi\sim\varphi+2\pi$.\\
The radial dependence of gauge fields $A$ and $C$ as well as their barred counterparts is fixed by
	\begin{subequations}\label{eq:ToyRadialDep}
	\begin{align}
		A(r,t,\varphi) & =b^{-1}(r)\left[a(t,\varphi)+\extd\,\right]b(r),\qquad C(r,t,\varphi)=\tilde{b}^{-1}(r)\left[c(t,\varphi)+\extd\,\right]\tilde{b}(r),\\
		\bar{A}(r,t,\varphi) & =b(r)\left[\bar{a}(t,\varphi)+\extd\,\right]b^{-1}(r),\qquad \bar{C}(r,t,\varphi)=\tilde{b}(r)\left[\bar{c}(t,\varphi)+\extd\,\right]\tilde{b}^{-1}(r),
	\end{align}	
	\end{subequations}
with
	\begin{equation}
		a(t,\varphi)=a_\varphi(t,\varphi)\extd\varphi+a_t(t,\varphi)\extd t\qquad\textrm{and}\qquad c(t,\varphi)=c_\varphi(t,\varphi)\extd\varphi+c_t(t,\varphi)\extd t,
	\end{equation}
and similarly for the barred part. In order to be able to straightforwardly take the limit of vanishing cosmological constant $\Lambda=-\frac{1}{\ell^2}\rightarrow0$ it is useful to choose the so called BMS gauge \cite{Barnich:2012aw} that in the Chern-Simons formulation is given by
    \begin{equation}
        b(r)=e^{\frac{r}{2\ell}\mathfrak{L}_{-1}}.
    \end{equation}
In addition we also set\footnote{It should be noted at this point that neither the choice of $b$ nor $\tilde{b}$ influences the canonical analysis or the asymptotic symmetries in any way. The choice of $b$ only influences a possible metric interpretation of the Chern-Simons connection \eqref{eq:TOYBCs}. However, since the gauge field does not modify the metric the specific choice of $\tilde{b}$ does not influence any results or physical interpretations in this work.} $\tilde{b}=e^{\frac{r}{2\ell}\mathfrak{S}}$.\\    
One can now choose boundary conditions as    
	\begin{subequations}\label{eq:TOYBCs}
	\begin{align}
		a_\varphi&=\mathfrak{L}_1-\frac{2\pi}{k}\left(\mathcal{L}-\frac{2\pi}{\kappa}\mathcal{K}^2\right)\mathfrak{L}_{-1},&
		a_t&=\mathfrak{L}_1-\frac{2\pi}{k}\left(\mathcal{L}-\frac{2\pi}{\kappa}\mathcal{K}^2\right)\mathfrak{L}_{-1},\\
		\bar{a}_\varphi&=\mathfrak{L}_1-\frac{2\pi}{k}\left(\bar{\mathcal{L}}-\frac{2\pi}{\bar{\kappa}}\bar{\mathcal{K}}^2\right)\mathfrak{L}_{-1},&
		\bar{a}_t&=-\mathfrak{L}_1+\frac{2\pi}{k}\left(\bar{\mathcal{L}}-\frac{2\pi}{\bar{\kappa}}\bar{\mathcal{K}}^2\right)\mathfrak{L}_{-1},\\
		c_\varphi&=\frac{4\pi}{\kappa}\mathcal{K}\mathfrak{S},&
		c_t&=\frac{4\pi}{\kappa}\mathcal{K}\mathfrak{S},\\
		\bar{c}_\varphi&=\frac{4\pi}{\bar{\kappa}}\bar{\mathcal{K}}\mathfrak{S},&
		\bar{c}_t&=-\frac{4\pi}{\bar{\kappa}}\bar{\mathcal{K}}\mathfrak{S},
	\end{align}
	\end{subequations}
where $\mathcal{L}\equiv\mathcal{L}(x^+)$, $\mathcal{K}\equiv\mathcal{K}(x^+)$, $\bar{\mathcal{L}}\equiv\bar{\mathcal{L}}(x^-)$, $\bar{\mathcal{K}}\equiv\bar{\mathcal{K}}(x^-)$ with $x^\pm=t\pm\varphi=\frac{u}{\ell}\pm\varphi$.\\
Since the main focus on this section is on the specific model in question and not the asymptotic symmetry analysis of the boundary conditions \eqref{eq:TOYBCs} we put all the technical details leading to the asymptotic symmetry algebra in Appendix~\ref{sec:CAToy}. As described in this appendix it is straightforward to show that the Fourier modes of the functions $\mathcal{L}$, $\mathcal{K}$ (as well as their barred counterparts) form the following asymptotic symmetry algebra:
	\begin{subequations}\label{eq:ToyVirU1}
	\begin{align}
		[\mathfrak{L}_n,\mathfrak{L}_m]&=(n-m)\mathfrak{L}_{n+m}+\frac{c}{12}n^3\delta_{n+m,0},\\
		[\mathfrak{L}_n,\mathfrak{K}_m]&=-m\mathfrak{K}_{n+m},\\
		[\mathfrak{K}_n,\mathfrak{K}_m]&=\frac{\kappa}{2}\,n\delta_{n+m,0},
	\end{align}
	\end{subequations}
where $c=6k$ (for the barred part we have $\bar{c}=c=6k$ and the $\mathfrak{u}(1)$ level is $\frac{\bar{\kappa}}{2}$).\\
The presence of the additional $\hat{\mathfrak{u}}(1)$ symmetries in \eqref{eq:ToyVirU1} gives rise to a one parameter automorphism of the algebra that is called ``spectral flow symmetry''. That means that one can define the following new generators (the same is true for the barred generators):
    \begin{equation}\label{eq:ToySpectralFlow}
        \tilde{\mathfrak{L}}_n:=\mathfrak{L}_n+u\mathfrak{K}_n+\frac{u^2}{4}\kappa\delta_{n,0},\qquad\tilde{\mathfrak{K}}:=\mathfrak{K}_n+u\frac{\kappa}{2}\delta_{n,0},
    \end{equation}
and that the new, tilded operators still satisfy the algebra \eqref{eq:ToyVirU1}. The presence of this automorphism also has some physical consequences. Since the spectral flow does not change the form of the algebra this means that also physical observables that depend on the underlying symmetries such as the thermal entropy of black hole solutions in this setup should also be invariant under the flow \eqref{eq:ToySpectralFlow}. In order to see this one can first determine the thermal entropy of the charged Ba{\~n}ados-Teitelboim-Zanelli (BTZ) black holes \cite{Banados:1992wn,Banados:1992gq} with mass $\ell M=\mathfrak{L}_0+\bar{\mathfrak{L}}_0$ and angular momentum $J=\mathfrak{L}_0-\bar{\mathfrak{L}}_0$ by employing
    \begin{equation}
        g_{\mu\nu}=\frac{\ell^2}{2}\left\langle(A_\mu-\bar{A}_\mu)(A_\nu-\bar{A}_\nu)\right\rangle,
    \end{equation}
and using the Bekenstein-Hawking area law, or by integrating the first law of (charged) black hole thermodynamics as outlined e.g. in Section~\ref{sec:FSCFirstLawThermalEntropy} for cosmological solutions in flat space. Both methods yield the same result, namely
    \begin{equation}\label{eq:ToyEntropy}
        S_{\textrm{Th}}=2\pi\sqrt{\frac{c}{6}\left(\mathfrak{L}_0-\frac{\mathfrak{K}_0^2}{\kappa}\right)}+2\pi\sqrt{\frac{\bar{c}}{6}\left(\bar{\mathfrak{L}}_0-\frac{\bar{\mathfrak{K}}_0^2}{\bar{\kappa}}\right)},
    \end{equation}
and it can be explicitly checked that, indeed, \eqref{eq:ToyEntropy} is invariant under the spectral flow \eqref{eq:ToySpectralFlow}.\\
Since we have written the boundary conditions \eqref{eq:TOYBCs} already in a form that makes a limit of vanishing cosmological constant easier to perform we will now briefly explain how the transition from AdS$_3$ to flat space works on the level of Chern-Simons connections\footnote{The limit also works in principle for the thermal entropy \eqref{eq:ToyEntropy}. However, there are some subtleties related to the relative sign between the two terms contributing to the entropy that one has to take care of as outlined in \cite{Riegler:2014bia,Fareghbal:2014qga} for the case of uncharged BTZ black holes.}. In order to do so we will use the so called Grassmann trick \cite{Krishnan:2013wta} that basically consists of replacing $\frac{1}{\ell}\rightarrow\epsilon$ and treating $\epsilon$ as a Grassmann parameter i.e. $\epsilon^2=0$. In order to take the limit it is first important to remember the relations between dualized spin connection $\omega$, dreibein $e$ and Chern-Simons connections $A$ and $\bar{A}$ as well as additional fields coming from the $\mathfrak{u}(1)$ gauge fields in AdS$_3$ i.e.
    \begin{equation}
        A=\omega+\epsilon e,\qquad \bar{A}=\omega-\epsilon e,\qquad C=\alpha+\epsilon \beta,\qquad \bar{C}=\alpha-\epsilon \beta.
    \end{equation}
The next step is to determine the components of the spin connection as well as the dreibein taking into account $\epsilon^2=0$ and then to use these components to write down new connections $\mathcal{A}$ and $\mathcal{C}$ as
    \begin{equation}
        \mathcal{A}=e^n \Mt_n+\omega^n\Lt_n,\qquad \mathcal{C}=\beta\Pt+\alpha\Jt,
    \end{equation}
where $\Mt_n$ and $\Lt_n$ are $\mathfrak{isl}(2,\mathbb{R})$ generators satisfying
    \begin{subequations}
    \begin{align}
        [\Lt_n,\Lt_m]&=(n-m)\Lt_{n+m},\\
        [\Lt_n,\Mt_m]&=(n-m)\Mt_{n+m},\\
        [\Mt_n,\Mt_m]&=0,
    \end{align}
    \end{subequations}
and $\Jt$ and $\Pt$ are $\mathfrak{u}(1)$ generators. In addition one can introduce the following new functions
    \begin{equation}
        \mathcal{L}=\frac{\mathcal{M}+\epsilon\mathcal{N}}{2\epsilon},\quad \mathcal{K}=\frac{\mathcal{P}+\epsilon\mathcal{J}}{2\epsilon},\quad \bar{\mathcal{L}}=\frac{\mathcal{M}-\epsilon\mathcal{N}}{2\epsilon},\quad \bar{\mathcal{K}}=\frac{\mathcal{P}-\epsilon\mathcal{J}}{2\epsilon},
    \end{equation}
as well as
    \begin{equation}
        \kappa=\frac{1}{2}\left(\frac{\kappa_P}{\epsilon}+\kappa_J\right),\qquad \bar{\kappa}=\frac{1}{2}\left(\frac{\kappa_P}{\epsilon}-\kappa_J\right),
    \end{equation}
and $k=\frac{\tilde{k}}{\epsilon}$. Using these functions, as well as the retarded time coordinate $u=\epsilon t$ one finds the following expressions for $\mathcal{A}$ and $\mathcal{C}$
    \begin{equation}
        \mathcal{A}=b^{-1}(a+\extd)b,\qquad \mathcal{C}=\tilde{b}^{-1}(c+\extd)\tilde{b},\qquad b=e^{\frac{r}{2}\Mt_{-1}},\qquad \tilde{b}=e^{\frac{r}{2}\Pt},
    \end{equation}
with    
	\begin{subequations}\label{eq:ToyFlatConnection}
	\begin{align}
		a_\varphi&=\Lt_1-\frac{\pi}{\tilde{k}}\left(\mathcal{M}-\frac{2\pi}{\kappa_P}\mathcal{P}^2\right)\Lt_{-1}-\frac{\pi}{\tilde{k}}\left(\mathcal{N}-\frac{4\pi}{\kappa_P}\mathcal{J}\mathcal{P}+2\pi\frac{\kappa_J}{\kappa_P^2}\mathcal{P}^2\right)\Mt_{-1},\\
		a_u&=\Mt_1-\frac{\pi}{\tilde{k}}\left(\mathcal{M}-\frac{2\pi}{\kappa_P}\mathcal{P}^2\right)\Mt_{-1},\\
		c_\varphi&=\frac{4\pi}{\kappa_P}\mathcal{P}\Jt+\frac{4\pi}{\kappa_P}\left(\mathcal{J}-\frac{\kappa_J}{\kappa_P}\mathcal{P}\right)\Pt,\\
		c_u&=\frac{4\pi}{\kappa_P}\mathcal{P}\Pt.
	\end{align}
	\end{subequations}	
We will see later in Section~\ref{sec:GravityModels} in more detail that the connection \eqref{eq:ToyFlatConnection}, indeed, gives rise to spectral flow symmetry in asymptotically flat spacetimes.

%--------------------------------------------------
\section{Spectral Flow Symmetry}\label{sec:SpectralFlow}
%--------------------------------------------------

The main purpose of this section is to introduce spectral flow symmetry as an automorphism of certain algebras that will be of interest in regards to the gravity models presented in this work. Initially we will motivate the most general form of these algebras as \.In\"on\"u--Wigner contractions of algebras that are relevant in the context of AdS$_3$/CFT$_2$ holography and exhibit spectral flow invariance. Later in Section \ref{sec:GravityModels} we then show how the resulting algebras can be alternatively motivated as asymptotic symmetry algebras of certain gravity models in 3D flat spacetimes.

%--------------------------------------------------
\subsection{$\mathfrak{bms}_3\inplus\hat{\mathfrak{u}}(1)\inplus\hat{\mathfrak{u}}(1)$}
%--------------------------------------------------

In \cite{Detournay:2016sfv} new asymptotic boundary conditions for flat spacetimes in 3D pure gravity\footnote{These boundary conditions can be seen as the flat space analogue of the Troessaert boundary conditions \cite{Troessaert:2013fma}.} were formulated that yield a symmetry algebra of the general form\footnote{For Einstein gravity one has $c_L=\kappa_J=0$ and $c_M=\frac{3}{G_N}$, $\kappa_P=-\frac{1}{4G_N}$, where $G_N$ is Newton's constant in 3D. For TMG in 3D flat space on the other hand one has $c_L=\frac{3}{\mu G_N}=-8\kappa_J$ and $c_M=\frac{3}{G_N}$, $\kappa_P=-\frac{1}{4G_N}$.}
	\begin{subequations}\label{eq:BMSU1Squared}
	\begin{align}
		[\Lt_n,\Lt_m]&=(n-m)\Lt_{n+m}+\frac{c_L}{12}n^3\delta_{n+m,0},\\
		[\Lt_n,\Mt_m]&=(n-m)\Mt_{n+m}+\frac{c_M}{12}n^3\delta_{n+m,0},\displaybreak[1]\\
		[\Lt_n,\Jt_m]&=-m\Jt_{n+m},\\
		[\Lt_n,\Pt_m]&=-m\Pt_{n+m},\\
		[\Mt_n,\Jt_m]&=-m\Pt_{n+m},\\
		[\Jt_n,\Jt_m]&=\frac{\kappa_J}{2}\,n\delta_{n+m,0},\\
		[\Jt_n,\Pt_m]&=\frac{\kappa_P}{2}\,n\delta_{n+m,0} \label{jp1},
	\end{align}
	\end{subequations}
where $n,m\in\mathbb{Z}$. This algebra can alternatively also be obtained as an \.In\"on\"u--Wigner contraction of two copies of a semi-direct product of a Virasoro algebra (with generators $\mathfrak{L}_n$ and $\bar{\mathfrak{L}}_n$) and an affine $\hat{\mathfrak{u}}(1)$ current algebra (with generators $\mathfrak{K}_n$ and $\bar{\mathfrak{K}}_n$) as in \eqref{eq:ToyVirU1}. It is straightforward to show that after the redefinitions
    \begin{subequations}
    \begin{align}
        \Lt_n:=&\mathfrak{L}_n-\bar{\mathfrak{L}}_{-n},&\Mt_n:=&\epsilon\left(\mathfrak{L}_n+\bar{\mathfrak{L}}_{-n}\right),\\
        \Jt_n:=&\mathfrak{K}_n-\bar{\mathfrak{K}}_{-n},&\Pt_n:=&\epsilon\left(\mathfrak{K}_n+\bar{\mathfrak{K}}_{-n}\right),\\
        c_L:=&c-\bar{c},&c_M:=&\epsilon\left(c+\bar{c}\right),\\
        \kappa_J:=&\kappa-\bar{\kappa},&\kappa_P:=&\epsilon\left(\kappa+\bar{\kappa}\right),        
    \end{align}
    \end{subequations}
one precisely obtains \eqref{eq:BMSU1Squared} in the limit $\epsilon\rightarrow0$.\\
Having an algebra like \eqref{eq:BMSU1Squared} where $\hat{\mathfrak{u}}(1)$ current algebras are present as subalgebras, a natural question to ask is if there is a spectral flow symmetry present. And indeed, for the algebra \eqref{eq:BMSU1Squared} it is straightforward to show that the following two-parameter flow preserves the general structure of the algebra i.e. the following transformation is an inner automorphism:
    \begin{subequations}\label{eq:BMSFlow}
    \begin{align}
        \tilde{\Lt}_n & :=\Lt_n+u\Jt_n+v\Pt_n+\frac{2uv\kappa_P+u^2\kappa_J}{4}\delta_{n,0},\\
        \tilde{\Mt}_n & :=\Mt_n+u\Pt_n+\frac{u^2}{4}\kappa_P\delta_{n,0},\\
        \tilde{\Jt}_n & :=\Jt_n+\frac{u\kappa_J+v\kappa_P}{2}\delta_{n,0},\\
        \tilde{\Pt}_n & :=\Pt_n+\frac{u}{2}\kappa_P\delta_{n,0},\\
    \end{align}
    \end{subequations}
where $u$ and $v$ are arbitrary real numbers.\\
At this point it is important to note that the fact that the algebra \eqref{eq:BMSU1Squared} allows for a two-parameter flow is closely linked to the fact that there are two a priori independent $\hat{\mathfrak{u}}(1)$ levels $\kappa_J$ and $\kappa_P$. As soon as one of the levels vanishes or they are no longer independent from each other one is left with only a single-parameter flow whose exact form can be deduced from \eqref{eq:BMSFlow}.

%--------------------------------------------------
\subsection{$\mathcal{N}=4$ Extended Super $\mathfrak{bms}_3$}
%--------------------------------------------------

As an additional example of an algebra exhibiting spectral flow invariance we now discuss an $\mathcal{N}=4$ supersymmetric extension of the $\mathfrak{bms}_3$ algebra that is given by the following non-vanishing relations:
    \begin{subequations} \label{n4subms}
    \begin{align}
        [\Lt_n,\Lt_m] & = (n-m)\Lt_{n+m}+\frac{c_L}{12}n(n^2-1)\delta_{n+m,0},\\
		[\Lt_n,\Mt_m] & = (n-m)\Mt_{n+m}+\frac{c_M}{12}n(n^2-1)\delta_{n+m,0},\\
		[\Lt_n,\Gt^\pm_r] & = (\tfrac{n}{2}-r)\Gt^\pm_{n+r},\\
		[\Lt_n,\Rt^\pm_r] & = (\tfrac{n}{2}-r)\Rt^\pm_{n+r},\\
		[\Mt_n,\Gt^\pm_r] & = (\tfrac{n}{2}-r)\Rt^\pm_{n+r},\\
		[\Lt_n,\Jt_m]&=-m\Jt_{n+m},\\
		[\Lt_n,\Pt_m]&=-m\Pt_{n+m},\\
		[\Mt_n,\Jt_m]&=-m\Pt_{n+m},\\
		[\Jt_n,\Gt^\pm_r] & = \pm \Gt^\pm_{n+r},\\ 
		[\Jt_n,\Rt^\pm_r] & = \pm \Rt^\pm_{n+r},\\ 
		[\Pt_n,\Gt^\pm_r] & = \pm \Rt^\pm_{n+r},\\
		\{\Gt^\pm_r,\Gt^\mp_s\} & = 2 \Lt_{r+s}\pm(r-s)\Jt_{r+s}+\frac{c_L}{3}\left(r^2-\tfrac{1}{4}\right)\delta_{r+s,0}, \label{gganticomm}\\
		\{\Gt^\pm_r,\Rt^\mp_s\} & = 2 \Mt_{r+s}\pm(r-s)\Pt_{r+s}+\frac{c_M}{3}\left(r^2-\tfrac{1}{4}\right)\delta_{r+s,0}, \label{granticomm}\\
		[\Jt_n,\Jt_m] & = \frac{c_L}{3}n\delta_{n+m,0},\\
		[\Jt_n,\Pt_m] & = \frac{c_M}{3}n\delta_{n+m,0} \label{jp2},
    \end{align}
    \end{subequations}
where $n,m\in\mathbb{Z}$ and $r,s\in\mathbb{Z}+\tfrac{1}{2}$ in the Neveu-Schwarz basis and $r,s\in\mathbb{Z}$ in the Ramond basis. The exact form of this algebra can be understood as taking the $\mathfrak{bms}_3$ algebra with its two possible central extensions and adding four supersymmetry generators in a certain way along with an $\hat{\mathfrak{u}}(1)\oplus\hat{\mathfrak{u}}(1)$ R-symmetry such that the (graded) Jacobi identities are satisfied\footnote{It should be noted that there is another possibility of extending the $\mathfrak{bms}_3$ algebra with $\mathcal{N}=4$ SUSY and $\hat{\mathfrak{u}}(1)\oplus\hat{\mathfrak{u}}(1)$ R-symmetry that has been described in \cite{Banerjee:2017gzj}.}.\\
Alternatively this algebra can be motivated by an \.In\"on\"u--Wigner contraction of two copies of the $\mathcal{N}=2$ superconformal algebra \cite{Ademollo:1975an}. Interestingly there are a number of different contractions of the super-conformal algebra resulting into non-unique yet sensible SUSY algebras on top of $\mathfrak{bms}_3$ \cite{deAzcarraga:2009ch,Sakaguchi:2009de,Mandal:2010gx,Banerjee:2016nio,Bagchi:2016yyf,Casali:2016atr}. The variations coming from different contractions are reflected mainly in the content of R-symmetry. Our chosen R-symmetry being fixed to be $\hat{\mathfrak{u}}(1) \oplus \hat{\mathfrak{u}}(1)$ can be seen as a ``despotic''\footnote{See \cite{Lodato:2016alv} for further details on this nomenclature.} contraction \cite{Mandal:2016lsa}  with generators $\mathfrak{L}_n$, $\bar{\mathfrak{L}}_n$, $\mathfrak{J}_n$, $\bar{\mathfrak{J}}_n$ as well as $\mathfrak{G}^\pm_n$, $\bar{\mathfrak{G}}^\pm_n$ and the central charges $c$ and $\bar{c}$. Then defining
    \begin{subequations}
    \begin{align}
        \Lt_n:=&\mathfrak{L}_n+\bar{\mathfrak{L}}_{n},&\Mt_n:=&\epsilon\left(\mathfrak{L}_n-\bar{\mathfrak{L}}_{n}\right),\\
        \Jt_n:=&\mathfrak{J}_n+\bar{\mathfrak{J}}_{n},&\Pt_n:=&\epsilon\left(\mathfrak{J}_n-\bar{\mathfrak{J}}_{n}\right),\\
        \Gt^\pm_n:=&\mathfrak{G}^\pm_n+\bar{\mathfrak{G}}^\pm_{n},&\Rt^\pm_n:=&\epsilon\left(\mathfrak{G}^\pm_n-\bar{\mathfrak{G}}^\pm_{n}\right),\\
        c_L:=&c+\bar{c},&c_M:=&\epsilon\left(c-\bar{c}\right),      
    \end{align}
    \end{subequations}
one obtains precisely\footnote{Please note that in \cite{Mandal:2016lsa} this algebra was called the $\mathcal{N}=4$ super Galilean conformal algebra.} \eqref{n4subms} in the limit $\epsilon\rightarrow0$.\\
As in the previous subsection one can check if the algebra \eqref{n4subms} is spectral flow invariant and, indeed, this is the case.\\
Introducing the discrete flows
    \begin{subequations}
    \begin{eqnarray}
        \tilde{\Gt}^{\pm}_r = \Gt^{\pm}_{r\pm u}, \\
        \tilde{\Rt}^{\pm}_r = \Rt^{\pm}_{r\pm u},
    \end{eqnarray}
    \end{subequations}
one finds that \eqref{gganticomm} and \eqref{granticomm} transform as
    \begin{subequations}
     \begin{eqnarray}
       && \{\tilde{\Gt}^\pm_r,\tilde{\Gt}^\mp_s\}  = 2 \tilde{\Lt}_{r+s}\pm(r-s)\tilde{\Jt}_{r+s}+\frac{c_L}{3}\left(r^2-\frac{1}{4}\right)\delta_{r+s,0},\\
		&&\{\tilde{\Gt}^\pm_r,\tilde{\Rt}^\mp_s\}  = 2 \tilde{\Mt}_{r+s}\pm(r-s)\tilde{\Pt}_{r+s}+\frac{c_M}{3}\left(r^2-\frac{1}{4}\right)\delta_{r+s,0},
     \end{eqnarray}
    \end{subequations}
where the transformation of the $\mathfrak{bms}_3$ generators and currents are:
    \begin{subequations} \label{reducedflow}
     \begin{eqnarray}
      &&\tilde{\Lt}_n = \Lt_n + u \Jt _n + \frac{c_L}{6} u^2 \delta_{n,0}, \\
      &&\tilde{\Mt}_n = \Mt_n + u \Pt_n + \frac{c_M}{6} u^2 \delta_{n,0} , \\
      &&\tilde{\Jt}_n = \Jt_n + \frac{c_L}{3} u \delta_{n,0}, \\
      &&\tilde{\Pt}_n = \Pt_n + \frac{c_M}{3} u \delta_{n,0}.
     \end{eqnarray}
    \end{subequations}
It is important to note here that, similar to the $\mathcal{N}=2$ superconformal algebra, the spectral flow \eqref{reducedflow} with $u=\frac{1}{2}$ provides an isomorphism between the Neveu-Schwarz basis and the Ramond basis of the algebra \eqref{n4subms}. In addition it is to be noted that the above flow \eqref{reducedflow} is a subset of the two-parameter flow encountered earlier in \eqref{eq:BMSFlow}. This can be seen by only taking the $u$-flow in \eqref{eq:BMSFlow} while freezing the $v$-flow and setting $ \kappa_P = \dfrac{2 c_{M}}{3}$ and $\kappa_J = \dfrac{2 c_{L}}{3} $. One should also take into account that this is natural, since \eqref{eq:BMSU1Squared} is just the bosonic subalgebra of \eqref{n4subms} with these identifications (compare \eqref{jp1} with \eqref{jp2}).

It is curious to note that although the algebra we just described \eqref{n4subms} has four SUSY generators, its R-symmetry is $\hat{\mathfrak{u}}(1) \oplus \hat{\mathfrak{u}}(1)$. One may wish to see if an extension of R-symmetry is allowed in this case so as to match that of the small or large $\mathcal{N}=4$ super-algebra encountered in super conformal field theories \cite{Ademollo:1975an, Sevrin:1988ew}. However such a possibility is ruled out as shown in \cite{Mandal:2016lsa}. On the other hand there are other ways in which $\mathcal{N}=4$ supersymmetry can be incorporated with $\mathfrak{bms}_3$ with greater amount of R-symmetry. We leave that topic for future explorations.

Motivated by the case of $\mathfrak{bms}_3$, a prescription of highest weight representation can be given observing that $\Lt_0, \Mt_0, \Jt_0, \Pt_0$ form a commuting set of generators. The highest weight states can be labelled by the eigenvalues of $\Lt_0$, $\Mt_0$, $\Jt_0$ and $\Pt_0$ as
    \begin{subequations}
    \begin{align}\label{eq:HilbertEigenvalues}
        \Lt_0|h_L,h_M,j,p\rangle & =h_L|h_L,h_M,j,p\rangle,& \Mt_0|h_L,h_M,j,p\rangle &=h_M|h_L,h_M,j,p\rangle,\\ \Jt_0|h_L,h_M,j,p\rangle & =j|h_L,h_M,j,p\rangle,& \Pt_0|h_L,h_M,j,p\rangle &=p|h_L,h_M,j,p\rangle.
    \end{align}
    \end{subequations}
In addition this state is subject to the condition of getting annihilated by all positive modes of the algebra.\\
With this representation, one can define a (anti-) chiral state like-wise in a SCFT via \footnote{Note that in the following, the state $|h_L,h_M,j,p\rangle$ does not need to be a primary or a highest weight state in the sense that it does not have to be annihilated by the positive modes of the algebra. But only satisfy \eqref{eq:HilbertEigenvalues}.} 
    \begin{equation}
    \Gt^+_{-1/2} |h_L,h_M,j,p\rangle =0 .
    \end{equation}
It is straightforward to see from the anti-commutators that the chiral state satisfies the bound $h_L = j/2$. Similarly anti-chiral primaries do satisfy $\Gt^-_{-1/2} |h_L,h_M,j,p\rangle =0$ and are restricted by the condition $h_L = - j/2$. All other states in the Hilbert space are bound by the (anti-) chiral states as for them $h_L \geq \frac{|j|}{2}$. This is based on the assumption that we have non-negative norm states in the spectrum. This is, indeed possible for certain values of the central extensions $c_L$, $c_M$, $\kappa_J$ and $\kappa_P$ as well as the weights $h_L$, $h_M$, $j$ and $p$. For that case this is a BPS shortening of the spectrum.\\
On the other hand such a shortening from the possible `R-chiral' states defined via $\Rt^+_{-1/2}|h_L,h_M,j,p\rangle =0$ is ruled out. This is because of the fact that the commuting generator $\Mt_0$ is not diagonalizable in the space of descendants generated by negative $\Lt_n$ modes. This is connected to non-unitarity of the highest weight representations of the $\mathfrak{bms}_3$ algebra \cite{Grumiller:2014lna,Riegler:2016hah}. %See the short digression in the footnote \footenotemark[\ref{footnotel}] for further clarification on this issue.

%--------------------------------------------------
\section{Gravity Models Exhibiting Spectral Flow Symmetry}\label{sec:GravityModels}
%--------------------------------------------------

In this section we present three different models of gravity in 3D with vanishing cosmological constant that exhibit spectral flow symmetry. Since this paper is focused on the algebraic and physical properties of spectral flow symmetry the presentation of the gravity models and the associated boundary conditions that lead to the desired boundary dynamics will be rather compact in the main body of the paper. For more details on the asymptotic analysis of said gravity models we refer the interested reader to Appendix~\ref{sec:CanonicalAnalysis}. 

%--------------------------------------------------
\subsection{Einstein Gravity}\label{sec:EinsteinFlow}
%--------------------------------------------------

Let us consider Einstein gravity with vanishing cosmological constant in 3D supplemented by Chern-Simons gauge fields $\mathcal{C}$ with coupling $\frac{\kappa_P}{4}$ written in a first order formalism \cite{Witten:1988hc} i.e.
    \begin{equation}\label{eq:EH+u1}
        I_{\textrm{EH}+\textrm{CS}}=\frac{k}{4\pi}\int R+\frac{\kappa_P}{16\pi}\int\left\langle \mathcal{C}\wedge\extd \mathcal{C}\right\rangle,
    \end{equation}
where $R$ is the Ricci scalar, $k=\frac{1}{4 G_N}$ with $G_N$ being Newton's constant in 3D and $\mathcal{C}$ a $\mathfrak{u}(1)\oplus\mathfrak{u}(1)$ valued gauge field and $\left\langle\ldots\right\rangle$ denotes a suitable invariant bilinear form on the gauge algebra of $\mathcal{C}$.\\
Up to boundary terms the action \eqref{eq:EH+u1} can be equivalently formulated in terms of an $\mathfrak{isl}(2,\mathbb{R})\oplus\mathfrak{u}(1)\oplus\mathfrak{u}(1)$ Chern-Simons theory with the following action
\begin{equation}\label{eq:CSFS}
		I_{\rm CS}[\mathcal{A}] = \frac{k}{4\pi} \int_{\mathcal{M}} \langle\mathcal{A} \wedge \extd \mathcal{A} +\frac23 \mathcal{A} \wedge \mathcal{A} \wedge \mathcal{A}\rangle\,+\frac{\kappa_P}{16\pi}\int_{\mathcal{M}}\left\langle \mathcal{C}\wedge\extd \mathcal{C}\right\rangle,
	\end{equation}
where $\mathcal{A}\in\mathfrak{isl}(2,\mathbb{R})$ and $\mathcal{C}\in\mathfrak{u}(1)\oplus\mathfrak{u}(1)$.	
We use a basis for $\mathfrak{isl}(2,\mathbb{R})\oplus\mathfrak{u}(1)\oplus\mathfrak{u}(1)$ with generators $\Lt_n,\,\Mt_n, \Jt$ and $\Pt$ with $n=0,\pm1$ that have the following non-vanishing Lie brackets:
	\begin{subequations}\label{eq:isl2RBasis}
	\begin{align}
		[\Lt_n,\Lt_m]&=(n-m)\Lt_{n+m},\\
		[\Lt_n,\Mt_n]&=(n-m)\Mt_{n+m}.
	\end{align}
	\end{subequations}
The corresponding invariant bilinear form is given by $\langle \Lt_n\Lt_m\rangle=\langle \Mt_n\Mt_m\rangle=\langle \Lt_n\Jt\rangle=\langle \Mt_n\Jt\rangle=\langle \Lt_n\Pt\rangle=\langle \Mt_n\Pt\rangle=\langle \Jt\Jt\rangle=\langle \Pt\Pt\rangle=0$ as well as
	\begin{subequations}\label{eq:ISLInvBilForm}
	\begin{align}
		\langle \Lt_n\Mt_m\rangle & =-2\left(
			\begin{array}{c|ccc}
				  &\Mt_1&\Mt_0&\Mt_{-1}\\
				\hline
				\Lt_1&0&0&1\\
				\Lt_0&0&-\frac{1}{2}&0\\
				\Lt_{-1}&1&0&0
			\end{array}\right),\\
		\langle \Jt\Pt\rangle & =2.	
	\end{align}		
	\end{subequations}
Let us assume that the topology of the manifold is that of a solid cylinder. In addition we choose coordinates such that there is a radial direction $0\leq r<\infty$ and the boundary of the cylinder is parametrized by a retarded time coordinate $-\infty<u<\infty$ as well as an angular coordinate $\varphi\sim\varphi+2\pi$.\\
We then fix the radial dependence of gauge fields $\mathcal{A}$ and $\mathcal{C}$ as
	\begin{equation}\label{eq:RadialDep}
		\mathcal{A}(r,u,\varphi)=b^{-1}(r)\left[a(u,\varphi)+\extd\,\right]b(r),\qquad \mathcal{C}(r,u,\varphi)=\tilde{b}^{-1}(r)\left[c(u,\varphi)+\extd\,\right]\tilde{b}(r),	
	\end{equation}
with
	\begin{equation}
		a(u,\varphi)=a_\varphi(u,\varphi)\extd\varphi+a_u(u,\varphi)\extd u,\qquad\textrm{and}\qquad c(u,\varphi)=c_\varphi(u,\varphi)\extd\varphi+c_u(u,\varphi)\extd u.
	\end{equation}	
A consequence of such a gauge choice is that the equations of motion $F=\extd\mathcal{A}+[\mathcal{A},\mathcal{A}]=0$ and $\bar{F}=\extd\mathcal{C}=0$ simplify drastically.\\
It is important to note that different choices of the group element $b$ will yield different geometrical interpretations. In order to interpret our boundary conditions as cosmological solutions later on we choose this group element as%\footnote{Since $\Mt_{-1}$ commutes with $\Jt$ and $\Pt$ this means that the gauge field $\mathcal{C}$ does not depend on $r$ and takes only values in the $\mathfrak{u}(1)\oplus \mathfrak{u}(1)$ part of the gauge algebra as intended.}
	\begin{equation}
		b(r)=e^{\frac{r}{2}\Mt_{-1}},
	\end{equation}
in addition we set $\tilde{b}=e^{\frac{r}{2}\Pt}$.\\	
After having completely specified our specific setup we are now ready to formulate boundary conditions. Using the inspiration coming from the flat limit of the AdS$_3$ example \eqref{eq:ToyFlatConnection} one can write down boundary conditions in terms of the gauge fields $a$ and $c$ as
	\begin{subequations}\label{eq:NewFSBCs}
	\begin{align}
		a_\varphi&=\Lt_1-\frac{\pi}{k}\left(\mathcal{M}-\frac{2\pi}{\kappa_P}\mathcal{P}^2\right)\Lt_{-1}-\frac{\pi}{k}\left(\mathcal{N}-\frac{4\pi}{\kappa_P}\mathcal{J}\mathcal{P}\right)\Mt_{-1},\\
		a_u&=\Mt_1-\frac{\pi}{k}\left(\mathcal{M}-\frac{2\pi}{\kappa_P}\mathcal{P}^2\right)\Mt_{-1},\\
		c_\varphi&=\frac{4\pi}{\kappa_P}\mathcal{P}\Jt+\frac{4\pi}{\kappa_P}\mathcal{J}\Pt,\\
		c_u&=\frac{4\pi}{\kappa_P}\mathcal{P}\Pt,
	\end{align}
	\end{subequations}
where the functions $\mathcal{M}$, $\mathcal{N}$, $\mathcal{J}$ and $\mathcal{P}$ are arbitrary functions of $u$ and $\varphi$.\\
Looking at the equations of motion $F=0$ and $\bar{F}=0$ one obtains very simple constraints on the (retarded) time evolution of the functions $\mathcal{M}$, $\mathcal{N}$, $\mathcal{J}$ and $\mathcal{P}$ as
	\begin{equation}\label{eq:EOM}
		\partial_u\mathcal{P}=\partial_u\mathcal{M}=0,\qquad\partial_u\mathcal{J}=\partial_\varphi\mathcal{P},\qquad\partial_u\mathcal{N}=\partial_\varphi\mathcal{M}.
	\end{equation}
That means that on-shell these functions can be written as
    \begin{subequations}\label{eq:StateOnShell}
    \begin{align}
    \mathcal{M} &=  \mathcal{M}(\varphi),& \mathcal{N} &= \mathcal{L}(\varphi) +u\mathcal{M}' \\
    \mathcal{P} &= \mathcal{P}(\varphi),&\mathcal{J} &= \mathcal{J}(\varphi) + u \mathcal{P}'.
    \end{align}
    \end{subequations}
After performing the canonical analysis presented in Appendix~\ref{sec:CAEinstein} one finds the following non-vanishing Dirac brackets for the state dependent functions:
	\begin{subequations}
	\begin{align}
	\{\mathcal{L}(\varphi),\mathcal{L}(\bar{\varphi})\}_{\textrm{D.B}}&=2\mathcal{L}\delta'-\delta\mathcal{L}',\\
	\{\mathcal{L}(\varphi),\mathcal{M}(\bar{\varphi})\}_{\textrm{D.B}}&=2\mathcal{M}\delta'-\delta\mathcal{M}'-\frac{k}{2\pi}\delta''',\\
	\{\mathcal{L}(\varphi),\mathcal{J}(\bar{\varphi})\}_{\textrm{D.B}}&=\mathcal{J}\delta'-\delta\mathcal{J}',\\
	\{\mathcal{L}(\varphi),\mathcal{P}(\bar{\varphi})\}_{\textrm{D.B}}&=\mathcal{P}\delta'-\delta\mathcal{P}',\\
	\{\mathcal{M}(\varphi),\mathcal{J}(\bar{\varphi})\}_{\textrm{D.B}}&=\mathcal{P}\delta'-\delta\mathcal{P}',\\
	\{\mathcal{J}(\varphi),\mathcal{P}(\bar{\varphi})\}_{\textrm{D.B}}&=\frac{\kappa_P}{4\pi}\delta',
	\end{align}
	\end{subequations}
where all functions appearing on the r.h.s are functions of $\bar{\varphi}$ and prime denotes differentiation with respect to the corresponding argument. Moreover ${\delta\equiv\delta(\varphi-\bar{\varphi})}$ and $\delta'\equiv\partial_\varphi\delta(\varphi-\bar{\varphi})$. Expanding the fields and delta distribution in terms of Fourier modes as
	\begin{subequations}\label{eq:EinsteinFourierModes}
	\begin{align}
		\mathcal{M}&=\frac{1}{2\pi}\sum\limits_{n\in\mathbb{Z}}\left(\Mt_n-\frac{k}{2}\delta_{n,0}\right)e^{-in\varphi},&
		\mathcal{L}&=\frac{1}{2\pi}\sum\limits_{n\in\mathbb{Z}}\Lt_ne^{-in\varphi},\\
		\mathcal{P}&=\frac{1}{2\pi}\sum\limits_{n\in\mathbb{Z}}\Pt_ne^{-in\varphi},&
		\mathcal{J}&=\frac{1}{2\pi}\sum\limits_{n\in\mathbb{Z}}\Jt_ne^{-in\varphi},\\
		\delta&=\frac{1}{2\pi}\sum\limits_{n\in\mathbb{Z}}e^{-in(\varphi-\bar{\varphi})},
	\end{align}
	\end{subequations}
and then replacing the Dirac brackets with commutators using $i\{\cdot,\cdot\}\rightarrow[\cdot,\cdot]$ one obtains the following non-vanishing commutation relations:
	\begin{subequations}\label{eq:ASAPrelim}
	\begin{align}
		[\Lt_n,\Lt_m]&=(n-m)\Lt_{n+m},\\
		[\Lt_n,\Mt_m]&=(n-m)\Mt_{n+m}+\frac{c_M}{12}n(n^2-1)\delta_{n+m,0},\\
		[\Lt_n,\Jt_m]&=-m\Jt_{n+m},\\
		[\Lt_n,\Pt_m]&=-m\Pt_{n+m},\\
		[\Mt_n,\Jt_m]&=-m\Pt_{n+m},\\
		[\Jt_n,\Pt_m]&=\frac{\kappa_P}{2}n\delta_{n+m,0},
	\end{align}
	\end{subequations}
which is exactly the algebra \eqref{eq:BMSU1Squared} with $c_L=\kappa_J=0$ and $c_M=12k$.

%--------------------------------------------------
\subsection{$\mathcal{N}=4$ Flat Supergravity}\label{sec:N4FlatSupergravity}
%--------------------------------------------------

In this subsection we show how one can obtain an asymptotic symmetry algebra of the form \eqref{n4subms} from an $\mathcal{N}=4$ flat Chern-Simons supergravity action\footnote{While finishing this work we became aware of the work by Oscar Fuentealba, Javier Matulich and Ricardo Troncoso \cite{Fuentealba:2017fck} that is also dealing with supergravity in three-dimensional asymptotically flat spacetimes that has a partial overlap with some of the things we consider in this paper. However, the main focus in their work is on a supergravity action that can be seen as a truncation of the ``democratic'' limit of $\mathcal{N}=(2,2)$ supergravity in contrast to the case treated in this work that corresponds to the ``despotic'' limit. Even though the theories considered differ slightly it is nice to see that physical observables like the thermal entropy of cosmological solutions match precisely and can be written as \eqref{eq:FlowInvEntropy}, as they should, since the bosonic part of both theories is the same.}.\\
Let us consider an action of the form
\begin{equation}\label{eq:SCSFS}
		I_{\rm CS}[\mathcal{A}] = \frac{k}{4\pi} \int_{\mathcal{M}} \langle\mathcal{A} \wedge \extd \mathcal{A} +\frac23 \mathcal{A} \wedge \mathcal{A} \wedge \mathcal{A}\rangle,
	\end{equation}
where $\mathcal{A}$ takes values in %$\mathfrak{isl}(2|2)$ \mnote{Correct?}\rnote{Not sure. Better to write `following Lie-superalgebra'} 
the Lie superalgebra given by the following non-vanishing commutation and anti-commutation relations
    \begin{subequations}\label{eq:isl21}
    \begin{align}
        [\Lt_n,\Lt_m] & = (n-m)\Lt_{n+m},\\
		[\Lt_n,\Mt_m] & = (n-m)\Mt_{n+m},\\
		[\Lt_n,\Gt^\pm_r] & = \left(\tfrac{n}{2}-r\right)\Gt^\pm_{n+r},\\
		[\Lt_n,\Rt^\pm_r] & = \left(\tfrac{n}{2}-r\right)\Rt^\pm_{n+r},\\
		[\Mt_n,\Gt^\pm_r] & = \left(\tfrac{n}{2}-r\right)\Rt^\pm_{n+r},\\
		[\Jt,\Gt^\pm_r] & = \pm \Gt^\pm_r,\\ 
		[\Jt,\Rt^\pm_r] & = \pm \Rt^\pm_r,\\ 
		[\Pt,\Gt^\pm_r] & = \pm \Rt^\pm_r,\\
		\{\Gt^\pm_r,\Gt^\mp_s\} & = 2 \Lt_{r+s}\pm(r-s)\Jt_{r+s},\\
		\{\Gt^\pm_r,\Rt^\mp_s\} & = 2 \Mt_{r+s}\pm(r-s)\Pt_{r+s},
    \end{align}
    \end{subequations}
where $n,m=0,\pm1$ and $r,s=\pm\frac{1}{2}$ that we call $\mathfrak{isl}(2|1)$\footnote{For more details on this nomenclature please refer to Appendix~\ref{sec:isl21}.}. The invariant bilinear form on this algebra is given by
	\begin{subequations}\label{eq:isl21InvBilinForm}
	\begin{align}
		\langle \Lt_n\,\Mt_m\rangle & =-2\left(
			\begin{array}{c|ccc}
				  &\Mt_1&\Mt_0&\Mt_{-1}\\
				\hline
				\Lt_1&0&0&1\\
				\Lt_0&0&-\frac{1}{2}&0\\
				\Lt_{-1}&1&0&0
			\end{array}\right) = \langle \Mt_n\,\Lt_m\rangle,
	\end{align}
	\begin{align}
		\langle \Gt^a_r\,\Rt^b_s\rangle & =4\left(
			\begin{array}{c|cccc}
                    &\Rt^+_\frac{1}{2}&\Rt^+_{-\frac{1}{2}}&\Rt^-_\frac{1}{2}&\Rt^-_{-\frac{1}{2}}\\
				\hline
				\Gt^+_\frac{1}{2} & 0 & 0 & 0 & -1 \\
				\Gt^+_{-\frac{1}{2}} & 0 & 0 & 1 & 0 \\
				\Gt^-_\frac{1}{2} & 0 & -1 & 0 & 0 \\
				\Gt^-_{-\frac{1}{2}} & 1 & 0 & 0 & 0
			\end{array}\right)=-\langle \Rt^a_r\,\Gt^b_s\rangle,
	\end{align}
	\begin{align}
		\langle \Jt_n\,\Pt_m\rangle & =-4 = \langle \Pt_n\,\Jt_m\rangle,
	\end{align}	
	\end{subequations}
and all other pairings of generators vanish.\\
We propose now the following boundary conditions:
    \begin{equation}
    \mathcal{A}(r,u,\varphi)=b^{-1}(r)\left[a(u,\varphi)+\extd\,\right]b(r),\qquad\textrm{and}\qquad b(r)=e^{\frac{r}{2}\Mt_{-1}},
    \end{equation}
and    
	\begin{subequations}\label{eq:NewSFSBCs}
	\begin{align}
		a_\varphi&=\Lt_1-\frac{\pi}{k}\left(\mathcal{M}+\frac{\pi}{4k}\mathcal{P}^2\right)\Lt_{-1}-\frac{\pi}{k}\left(\mathcal{N}+\frac{\pi}{2k}\mathcal{J}\mathcal{P}\right)\Mt_{-1}-\frac{\pi}{2k}\mathcal{P}\Jt-\frac{\pi}{2k}\mathcal{J}\Pt\nonumber\\
		    &\quad -\frac{\pi}{2k}\mathcal{R}^-\Gt^+_{-\frac{1}{2}}-\frac{\pi}{2k}\mathcal{R}^+\Gt^-_{-\frac{1}{2}}-\frac{\pi}{2k}\mathcal{G}^-\Rt^+_{-\frac{1}{2}}-\frac{\pi}{2k}\mathcal{G}^+\Rt^-_{-\frac{1}{2}},\\
		a_u&=\Mt_1-\frac{\pi}{k}\left(\mathcal{M}+\frac{\pi}{4k}\mathcal{P}^2\right)\Mt_{-1}-\frac{\pi}{2k}\mathcal{P}\Pt-\frac{\pi}{2k}\mathcal{R}^-\Rt^+_{-\frac{1}{2}}-\frac{\pi}{2k}\mathcal{R}^+\Rt^-_{-\frac{1}{2}},
	\end{align}
	\end{subequations}
where $\mathcal{M}$, $\mathcal{N}$, $\mathcal{J}$, $\mathcal{P}$ are commuting functions of $u$ and $\varphi$ and $\mathcal{R}^a$ and $\mathcal{G}^a$ are anticommuting Grassmann valued functions of $u$ and $\varphi$.\\
As in the Einstein gravity case, the equations of motion $F=\extd\mathcal{A}+\mathcal{A}\wedge\mathcal{A}=0$ put certain constraints on the functions appearing in \eqref{eq:NewSFSBCs} as
	\begin{equation}\label{eq:SuperEOM}
		\partial_u\mathcal{P}=\partial_u\mathcal{M}=\partial_u\mathcal{G}^\pm=\partial_u\mathcal{R}^\pm=0,\qquad\partial_u\mathcal{J}=\partial_\varphi\mathcal{P},\qquad\partial_u\mathcal{N}=\partial_\varphi\mathcal{M},\qquad\partial_u\mathcal{G}^\pm=\partial_\varphi\mathcal{R}^\pm.
	\end{equation}
This means that on-shell these functions can be written as
    \begin{subequations}\label{eq:SuperStateOnShell}
    \begin{align}
    \mathcal{M} &=  \mathcal{M}(\varphi),& \mathcal{N} &= \mathcal{L}(\varphi) +u\mathcal{M}' \\
    \mathcal{R}^\pm &=  \mathcal{R}^\pm(\varphi),& \mathcal{G}^\pm &= \mathcal{G}^\pm(\varphi) +u\left(\mathcal{R}^\pm\right)',\\
    \mathcal{P} &= \mathcal{P}(\varphi),&\mathcal{J} &= \mathcal{J}(\varphi) + u \mathcal{P}'.    
    \end{align}
    \end{subequations}
After determining the boundary condition preserving gauge transformations and the corresponding canonical boundary charges as shown in Appendix~\ref{sec:CASuper} one finds the following non-vanishing Dirac brackets for the state dependent functions:
	\begin{subequations}
	\begin{align}
	\{\mathcal{L}(\varphi),\mathcal{L}(\bar{\varphi})\}_{\textrm{D.B}}&=2\mathcal{L}\delta'-\delta\mathcal{L}',\\
	\{\mathcal{L}(\varphi),\mathcal{M}(\bar{\varphi})\}_{\textrm{D.B}}&=2\mathcal{M}\delta'-\delta\mathcal{M}'-\frac{k}{2\pi}\delta''',\\
    \{\mathcal{L}(\varphi),\mathcal{G}^\pm(\bar{\varphi})\}_{\textrm{D.B}}&=\frac{3}{2}\mathcal{G}^\pm\delta'-\delta(\mathcal{G}^\pm)',\\
    \{\mathcal{L}(\varphi),\mathcal{R}^\pm(\bar{\varphi})\}_{\textrm{D.B}}&=\frac{3}{2}\mathcal{R}^\pm\delta'-\delta(\mathcal{R}^\pm)',\\
    \{\mathcal{M}(\varphi),\mathcal{G}^\pm(\bar{\varphi})\}_{\textrm{D.B}}&=\frac{3}{2}\mathcal{R}^\pm\delta'-\delta(\mathcal{R}^\pm)',\\
	\{\mathcal{L}(\varphi),\mathcal{J}(\bar{\varphi})\}_{\textrm{D.B}}&=\mathcal{J}\delta'-\delta\mathcal{J}',\\
	\{\mathcal{L}(\varphi),\mathcal{P}(\bar{\varphi})\}_{\textrm{D.B}}&=\mathcal{P}\delta'-\delta\mathcal{P}',\\
	\{\mathcal{M}(\varphi),\mathcal{J}(\bar{\varphi})\}_{\textrm{D.B}}&=\mathcal{P}\delta'-\delta\mathcal{P}',\\
	\{\mathcal{J}(\varphi),\mathcal{G}^\pm(\bar{\varphi})\}_{\textrm{D.B}}&=\mp\mathcal{G}^\pm\delta,\\
	\{\mathcal{J}(\varphi),\mathcal{R}^\pm(\bar{\varphi})\}_{\textrm{D.B}}&=\mp\mathcal{R}^\pm\delta,\\
	\{\mathcal{P}(\varphi),\mathcal{G}^\pm(\bar{\varphi})\}_{\textrm{D.B}}&=\mp\mathcal{G}^\pm\delta,\\
	\{\mathcal{G}^\pm(\varphi),\mathcal{G}^\mp(\bar{\varphi})\}_{\textrm{D.B}}&=2\mathcal{L}\delta\pm\delta\mathcal{J}'\mp2\mathcal{J}\delta',\label{eq:Anti1}\\
	\{\mathcal{G}^\pm(\varphi),\mathcal{R}^\mp(\bar{\varphi})\}_{\textrm{D.B}}&=2\mathcal{M}\delta\pm\delta\mathcal{P}'\mp2\mathcal{P}\delta'-\frac{2}{k\pi}\delta'',\label{eq:Anti2}\\
	\{\mathcal{J}(\varphi),\mathcal{P}(\bar{\varphi})\}_{\textrm{D.B}}&=-\frac{2k}{\pi}\delta',
	\end{align}
	\end{subequations}
where all functions appearing on the r.h.s are functions of $\bar{\varphi}$ and prime denotes differentiation with respect to the corresponding argument. Moreover ${\delta\equiv\delta(\varphi-\bar{\varphi})}$ and $\delta'\equiv\partial_\varphi\delta(\varphi-\bar{\varphi})$. Expanding the fields and delta distribution\footnote{For \eqref{eq:Anti1} and \eqref{eq:Anti2} we used $\delta=\frac{1}{2\pi}\sum\limits_{n\in\mathbb{Z}+\frac{1}{2}}e^{-in(\varphi-\bar{\varphi})}$. For more details on this please refer to e.g. \cite{Riegler:2012fa}.} in terms of Fourier modes as
	\begin{subequations}\label{eq:SuperEinsteinFourierModes}
	\begin{align}
		\mathcal{M}&=\frac{1}{2\pi}\sum\limits_{n\in\mathbb{Z}}\left(\Mt_n-\frac{k}{2}\delta_{n,0}\right)e^{-in\varphi},&
		\mathcal{L}&=\frac{1}{2\pi}\sum\limits_{n\in\mathbb{Z}}\Lt_ne^{-in\varphi},\\
		\mathcal{R}^\pm&=\frac{1}{2\pi}\sum\limits_{n\in\mathbb{Z}+\frac{1}{2}}\Rt^\pm_ne^{-in\varphi},&
		\mathcal{G}^\pm&=\frac{1}{2\pi}\sum\limits_{n\in\mathbb{Z}+\frac{1}{2}}\Gt^\pm_ne^{-in\varphi},\\		
		\mathcal{P}&=-\frac{i}{2\pi}\sum\limits_{n\in\mathbb{Z}}\Pt_ne^{-in\varphi},&
		\mathcal{J}&=-\frac{i}{2\pi}\sum\limits_{n\in\mathbb{Z}}\Jt_ne^{-in\varphi},\\
		\delta&=\frac{1}{2\pi}\sum\limits_{n\in\mathbb{Z}}e^{-in(\varphi-\bar{\varphi})},
	\end{align}
	\end{subequations}
and then replacing the Dirac brackets with commutators using $i\{\cdot,\cdot\}_{\textrm{D.B}}\rightarrow[\cdot,\cdot]$ for the bosonic fields and with anticommutators using $\{\cdot,\cdot\}_{\textrm{D.B}}\rightarrow\{\cdot,\cdot\}$ for the fermionic fields one obtains the following non-vanishing commutation relations:    
    \begin{subequations} \label{n4submsSUGRA}
    \begin{align}
        [\Lt_n,\Lt_m] & = (n-m)\Lt_{n+m},\\
		[\Lt_n,\Mt_m] & = (n-m)\Mt_{n+m}+\frac{c_M}{12}n(n^2-1)\delta_{n+m,0},\\
		[\Lt_n,\Gt^\pm_r] & = \left(\tfrac{n}{2}-r\right)\Gt^\pm_{n+r},\\
		[\Lt_n,\Rt^\pm_r] & = \left(\tfrac{n}{2}-r\right)\Rt^\pm_{n+r},\\
		[\Mt_n,\Gt^\pm_r] & = \left(\tfrac{n}{2}-r\right)\Rt^\pm_{n+r},\\
		[\Lt_n,\Jt_m]&=-m\Jt_{n+m},\\
		[\Lt_n,\Pt_m]&=-m\Pt_{n+m},\\
		[\Mt_n,\Jt_m]&=-m\Pt_{n+m},\\
		[\Jt_n,\Gt^\pm_r] & = \pm \Gt^\pm_{n+r},\\ 
		[\Jt_n,\Rt^\pm_r] & = \pm \Rt^\pm_{n+r},\\ 
		[\Pt_n,\Gt^\pm_r] & = \pm \Rt^\pm_{n+r},\\
		\{\Gt^\pm_r,\Gt^\mp_s\} & = 2 \Lt_{r+s}\pm(r-s)\Jt_{r+s},\\
		\{\Gt^\pm_r,\Rt^\mp_s\} & = 2 \Mt_{r+s}\pm(r-s)\Pt_{r+s}+\frac{c_M}{3}\left(r^2-\frac{1}{4}\right)\delta_{r+s,0},\\
		[\Jt_n,\Pt_m] & = \frac{c_M}{3}n\delta_{n+m,0},
    \end{align}
    \end{subequations}
with $c_M=12k$.\\
Comparison with the algebra \eqref{n4subms} shows that this asymptotic symmetry algebra is, indeed, a special case with $\kappa_P=\frac{2c_M}{3}$ and $c_L=\kappa_J=0$ of the more general algebra \eqref{n4subms}.

%--------------------------------------------------
\subsection{Einstein Gravity and $\mathcal{N}=4$ Supergravity ``Reloaded''}\label{sec:Reloaded}
%--------------------------------------------------

The two models presented previously were able to realize subsets with $c_L=0$ and $\kappa_J=0$ of the general algebras \eqref{eq:BMSU1Squared} and \eqref{n4subms} as their asymptotic symmetry algebras. As such it would be nice to also have examples of models that realize the full algebras with all central extensions being non-zero. In the Chern-Simons formulation of gravity there is one particular efficient way of achieving this, that is, by changing the invariant bilinear form used in the Chern-Simons action such as in e.g. \cite{Giacomini:2006dr,Barnich:2014cwa} in order to obtain ``reloaded'' versions of Einstein und supergravity theories. We will first start with the Einstein gravity case treated in Sec.~\ref{sec:EinsteinFlow} and explain the main features of this construction before extending these considerations to the $\mathcal{N}=4$ supergravity story.\\
Consider a model whose Chern-Simons action has the same general form as in \eqref{eq:CSFS} with a bilinear form that satisfies \eqref{eq:ISLInvBilForm} as well as in addition
    \begin{equation}\label{eq:InvBilFormModEinstein}
        \langle \Lt_n\Lt_m\rangle = \mu \langle \Lt_n\Mt_m\rangle,\qquad \langle \Jt_n\Jt_m\rangle = \nu \langle \Jt_n\Pt_m\rangle.
    \end{equation}
This change does not directly influence the canonical analysis performed for the Einstein gravity case and the boundary conditions \eqref{eq:NewFSBCs}. Hence also the boundary condition preserving gauge transformations are still given by \eqref{eq:BCPGTs}. However, the resulting canonical charges and thus also the asymptotic symmetry algebra are affected by this change of invariant bilinear form. Thus one obtains the following variation of the canonical boundary charge:
    \begin{equation}
        \delta Q=\int\extd\varphi\left(\epsilon_\mathcal{L}\left(\delta\mathcal{L}+\mu\delta\mathcal{M}-\frac{4\pi}{\kappa_P}(\mu-\nu)\delta\mathcal{P}\mathcal{P}\right)+\epsilon_\mathcal{M}\delta\mathcal{M}+\epsilon_\mathcal{J}(\delta\mathcal{J}+\nu\delta\mathcal{P})+\epsilon_\mathcal{P}\delta\mathcal{P}\right).
    \end{equation}
One can now define new functions $\tilde{\mathcal{L}}$ and $\tilde{\mathcal{J}}$ as
    \begin{equation}\label{eq:ReloadedEinsteinCharges}
        \tilde{\mathcal{L}}:=\mathcal{L}+\mu\mathcal{M}-\frac{2\pi}{\kappa_P}(\mu-\nu)\mathcal{P}^2,\qquad
        \tilde{\mathcal{J}}:=\mathcal{J}+\nu\mathcal{P},
    \end{equation}
such that the variation of the canonical charge simplifies to
    \begin{equation}
        \delta Q = \int\extd\varphi\left(\epsilon_\mathcal{L}\delta\tilde{\mathcal{L}}+\epsilon_\mathcal{M}\delta\mathcal{M}+\epsilon_\mathcal{J}\delta\tilde{\mathcal{J}}+\epsilon_\mathcal{P}\delta\mathcal{P}\right).
    \end{equation}
It is then straightforward to show that the Fourier modes of the functions $\tilde{\mathcal{L}}$, $\mathcal{M}$, $\tilde{\mathcal{J}}$, $\mathcal{P}$ satisfy the algebra \eqref{eq:BMSU1Squared} with
    \begin{equation}
        c_L=12\mu k,\qquad c_M=12k,\qquad \kappa_J=\nu \kappa_P.
    \end{equation}
Similarly one can take a model with a Chern-Simons form \eqref{eq:SCSFS} and an invariant bilinear form that satisfies \eqref{eq:isl21InvBilinForm} as well as in addition
    \begin{equation}\label{eq:InvBilFormModSUSY}
        \langle \Lt_n\Lt_m\rangle = \mu \langle \Lt_n\Mt_m\rangle,\qquad \langle \Gt^a_r\,\Gt^b_s\rangle=\mu\langle \Gt^a_r\,\Rt^b_s\rangle,\qquad\langle \Jt_n\Jt_m\rangle = \mu \langle \Jt_n\Pt_m\rangle,
    \end{equation}
and use the same arguments as before to arrive at the following variation of the canonical boundary charges:  
    \begin{equation}\label{eq:SuperCanChargesVariationReloaded}
        \delta Q[\epsilon]=\int\extd\varphi\left(\epsilon_\mathcal{L}\delta\tilde{\mathcal{L}}+\epsilon_\mathcal{M}\delta\mathcal{M}+\epsilon_\mathcal{G}^+\delta\tilde{\mathcal{G}}^++\epsilon_\mathcal{G}^-\delta\tilde{\mathcal{G}}^-+\epsilon_\mathcal{R}^+\delta\mathcal{R}^++\epsilon_\mathcal{R}^-\delta\mathcal{R}^-+\epsilon_\mathcal{J}\delta\tilde{\mathcal{J}}+\epsilon_\mathcal{P}\delta\mathcal{P}\right),
    \end{equation}
where
    \begin{equation}\label{eq:ReloadedSUSYCharges}
        \tilde{\mathcal{L}}:=\mathcal{L}+\mu\mathcal{M},\qquad
        \tilde{\mathcal{G}}^\pm:=\mathcal{G}^\pm+\mu \mathcal{R}^\pm,\qquad
        \tilde{\mathcal{J}}:=\mathcal{J}+\nu\mathcal{P}.
    \end{equation}
One can again straightforwardly verify that the asymptotic symmetry algebra spanned by the Fourier modes of the functions appearing in the variation of the canonical boundary charge \eqref{eq:SuperCanChargesVariationReloaded} is given by \eqref{n4subms} with $c_L=12\mu k$ and $c_M=12k$.

%--------------------------------------------------
\section{Thermal Entropy: Gravity Side} \label{sec:ThermalGravity}
%--------------------------------------------------

In the previous Section~\ref{sec:GravityModels} we presented different gravity models and associated boundary conditions that lead to asymptotic symmetry algebras of the form presented in Section~\ref{sec:SpectralFlow}. In this section we compute the thermal entropy of cosmological solutions in these models and show that the resulting expressions are invariant under (a subset of) the spectral flow \eqref{eq:BMSFlow}. This is an important check to our claim that we have, indeed, found gravity models that exhibit a non-trivial spectral flow symmetry. The necessity for such a check arises from cases such as e.g. the one presented in \cite{Detournay:2016sfv} where the asymptotic symmetry algebra itself is in principle spectral flow invariant, but physical observables such as the thermal entropy of cosmological solutions are only trivially spectral flow invariant.\\
We determine the thermal entropy in two ways. Since it might be more intuitive for readers used to the second order formulation of gravity to compute it via the horizon area of flat space cosmologies we will start by doing exactly this. The other way might be more accessible for readers used to the Chern-Simons formulation of gravity that involves functionally integrating the first law of flat space cosmologies as well as imposing conditions on the holonomies of the Chern-Simons connection.

%--------------------------------------------------
\subsection{Horizon Area}
%--------------------------------------------------

We will begin this subsection by first determining the thermal entropy for the Einstein gravity case described in Section~\ref{sec:EinsteinFlow} as well as the $\mathcal{N}=4$ flat supergravity case described in Section~\ref{sec:N4FlatSupergravity}. To this end we first have to translate the boundary conditions \eqref{eq:NewFSBCs} and \eqref{eq:NewSFSBCs} to a metric form. This can be done by extracting the dreibein $e$ from the Chern-Simons connection $\mathcal{A}$ via \cite{Witten:1988hc}
	\begin{equation}
		\mathcal{A}=\omega^a\Lt_a+e^a\Mt_a,
	\end{equation}
for $a=0,\pm1$. Then using
	\begin{equation}
		\eta_{ab}=-2\left(
			\begin{array}{c|ccc}
				  &\Mt_1&\Mt_0&\Mt_{-1}\\
				\hline
				\Mt_1&0&0&1\\
				\Mt_0&0&-\frac{1}{2}&0\\
				\Mt_{-1}&1&0&0
			\end{array}\right),
	\end{equation}
one can recover a metric formulation via 
	\begin{equation}
		g_{\mu\nu}=\eta_{ab}e^a_\mu e^b_\nu.
	\end{equation}
For the boundary conditions \eqref{eq:NewFSBCs} as well as \eqref{eq:NewSFSBCs} this leads to the following metric:
	\begin{align}\label{eq:AdSBMSMetricNew}
		\extd s^2&=\frac{4\pi}{k}\left(\mathcal{M}-\frac{2\pi}{\kappa_P}\mathcal{P}^2\right)\extd u^2+\frac{4\pi}{k}\left(\mathcal{N}-\frac{4\pi}{\kappa_P}\mathcal{J}\mathcal{P}\right)\extd u\extd\varphi-2\extd r\extd u+r^2\extd\varphi^2,
	\end{align}
where one has $\kappa_P=-8k$ for the $\mathcal{N}=4$ supersymmetric boundary conditions \eqref{eq:NewSFSBCs} and $\kappa_P$ is arbitrary for the Einstein gravity case \eqref{eq:NewFSBCs}.\\ 	
In order to compute the area of the horizon one first has to locate the event horizon by looking at the point where the determinant of the induced metric on slices of constant radius vanishes, that is
    \begin{equation}
        g_{uu}\,g_{\varphi\varphi}-\left(g_{u\varphi}\right)^2=0.
    \end{equation}
For the metric \eqref{eq:AdSBMSMetricNew} this happens at
    \begin{equation}\label{eq:Horizon}
        r_{\rm H}=\frac{|\mathcal{N}-\frac{4\pi}{\kappa_P}\mathcal{J}\mathcal{P}|}{\sqrt{\frac{k}{\pi}\left(\mathcal{M}-\frac{2\pi}{\kappa_P}\mathcal{P}^2\right)}}.
    \end{equation}
One can now compute the area of the horizon by
    \begin{equation}\label{eq:HorizonArea}
        {\rm A}_{\rm H}=\int\limits_0^{2\pi}\extd\varphi\sqrt{|g_{\varphi\varphi}|}\,\Big|_{r=r_{\rm H}}=\int\limits_0^{2\pi}\extd\varphi\frac{|\mathcal{N}-\frac{4\pi}{\kappa_P}\mathcal{J}\mathcal{P}|}{\sqrt{\frac{k}{\pi}\left(\mathcal{M}-\frac{2\pi}{\kappa_P}\mathcal{P}^2\right)}}.
    \end{equation}
For the zero mode solutions\footnote{Note that in the flat supergravity case the zero modes read $\mathcal{N}=\frac{1}{2\pi}\Lt_0$, $\mathcal{M}=\frac{1}{2\pi}\Mt_0$,	$\mathcal{J}=\frac{i}{2\pi}\Jt_0$ and $\mathcal{P}=\frac{i}{2\pi}\Pt_0$.} $\mathcal{N}=\frac{1}{2\pi}\Lt_0$, $\mathcal{M}=\frac{1}{2\pi}\Mt_0$,	$\mathcal{J}=\frac{1}{2\pi}\Jt_0$ and $\mathcal{P}=\frac{1}{2\pi}\Pt_0$ this can be trivially integrated to yield the area
    \begin{equation}
        {\rm A}_{\rm H}^0=2\pi\frac{|\Lt_0-2\frac{\Jt_0\Pt_0}{\kappa_P}|}{\sqrt{2k\left(\Mt_0-\frac{\Pt_0^2}{\kappa_P}\right)}},
    \end{equation}
and the corresponding thermal entropy
    \begin{equation}\label{eq:ThermalEntropyArea}
        S_{\rm Th}=\frac{{\rm A}_{\rm H}^0}{4G_N}=\frac{\pi}{6}\frac{c_M\left(\Lt_0-2\frac{\Jt_0\Pt_0}{\kappa_P}\right)}{\sqrt{\frac{c_M}{6}\left(\Mt_0-\frac{\Pt_0^2}{\kappa_P}\right)}},
    \end{equation}
with $c_M=12k$ and $\kappa_P=8k$ for the $\mathcal{N}=4$ supersymmetric boundary conditions \eqref{eq:NewSFSBCs} and $\kappa_P=$arbitrary for the Einstein gravity case \eqref{eq:NewFSBCs}. It is straightforward to check that this entropy formula is, indeed, invariant under the spectral flow \eqref{eq:BMSFlow}.\\ 
At this point it is interesting to compare the results for the thermal entropy in the Einstein gravity case \eqref{eq:ThermalEntropyArea} with the thermal entropy found in \cite{Detournay:2016sfv}. Even though the asymptotic symmetry algebras are both subsets of the general form \eqref{eq:BMSU1Squared} and are invariant under spectral flow symmetry, the thermal entropy for cosmological solutions found in \cite{Detournay:2016sfv} is only trivially spectral flow invariant. The main reason for that is basically that, in the cases treated in this paper the horizon of the cosmological solutions get modified in the presence of the additional $\hat{\mathfrak{u}}(1)$ symmetries whereas this does not happen in the model considered in \cite{Detournay:2016sfv}. 

%--------------------------------------------------
\subsection{First Law of Flat Space Cosmologies}\label{sec:FSCFirstLawThermalEntropy}
%--------------------------------------------------

The main goal of this subsection is to determine the thermal entropy in a slightly different way than in the previous subsection. That is, we argue that the first law of (charged) flat space cosmologies is equivalent to imposing conditions on the holonomies of the Chern-Simons connections in question. The resulting entropy after integrating that first law then yields exactly the same expression for the thermal entropy as determined previously using the area law in \eqref{eq:ThermalEntropyArea}. Since there are small technical but no conceptual differences in the way one determines mass, angular momentum, the electric charges and their corresponding chemical potentials for the Einstein gravity, $\mathcal{N}=4$ flat supergravity case as well as their ``reloaded'' versions we will first present the Einstein case in more detail and then briefly comment on the other two cases.\\
In order to determine the Entropy of a flat space cosmology in 3D flat space Einstein gravity described by the connection \eqref{eq:NewFSBCs} we make use of the first law of flat space cosmologies \cite{Bagchi:2012xr} with inverse temperature $\beta_T$, mass $M$, angular velocity $\Omega$ and angular momentum $J$. Since for the case at hand one has additional electric potentials $\Phi_\mathcal{J}$ and $\Phi_\mathcal{P}$ and their associated electric charges $Q_\mathcal{J}$ and $Q_\mathcal{P}$ one expects the first law to account for those additional charges and potentials as
    \begin{equation}\label{eq:FirstLaw}
        \delta M=-T\delta S_{\textrm{Th}}+\Omega\delta J-\Phi_\mathcal{J}\delta Q_\mathcal{J}-\Phi_\mathcal{P}\delta Q_\mathcal{P}.
    \end{equation}
In a metric formulation the mass and angular momentum are associated to the charges of the global asymptotic Killing vectors (AKVs) $\partial_u$ and $\partial_\varphi$ respectively. Using that the gauge parameters \eqref{eq:GaugeParameter} that preserve the boundary conditions \eqref{eq:NewFSBCs} are related to the AKVs via $\epsilon+\bar{\epsilon}=\xi^\mu(\mathcal{A}_\mu+\mathcal{C}_\mu)$ (see e.g. \cite{Henneaux:1984ji,Bunster:2014cna,Perez:2015jxn}) one can determine (the variation of) mass and angular momentum of the cosmological solutions \eqref{eq:NewFSBCs} via
    \begin{subequations}\label{eq:VariationMassAngMomFormula}
    \begin{align}
        \delta M &:= \delta Q[\epsilon|_{\partial_u}]+\delta Q[\bar{\epsilon}|_{\partial_u}]=\frac{k}{2\pi}\int\extd\varphi\langle\mathcal{A}_u\delta\mathcal{A}_\varphi\rangle+\frac{\kappa_P}{8\pi}\int\extd\varphi\langle\mathcal{C}_u\delta\mathcal{C}_\varphi\rangle,\\
        \delta J &:= \delta Q[\epsilon|_{\partial_\varphi}]+\delta Q[\bar{\epsilon}|_{\partial_\varphi}]=\frac{k}{2\pi}\int\extd\varphi\langle\mathcal{A}_\varphi\delta\mathcal{A}_\varphi\rangle+\frac{\kappa_P}{8\pi}\int\extd\varphi\langle\mathcal{C}_\varphi\delta\mathcal{C}_\varphi\rangle.
    \end{align}
    \end{subequations}
Plugging in the expression for the connection \eqref{eq:NewFSBCs} one sees that only the zero modes of $\mathcal{N}(\varphi)$, $\mathcal{M}(\varphi)$, $\mathcal{J}(\varphi)$ and $\mathcal{P}(\varphi)$ that we again denote by $\Lt_0$, $\Mt_0$, $\Jt_0$ and $\Pt_0$ respectively contribute in \eqref{eq:VariationMassAngMomFormula} as
    \begin{equation}
        \delta M = \delta\Mt_0,\qquad \delta J =  \delta \Lt_0.
    \end{equation}
The electric charges are determined from $\delta Q[\bar{\epsilon}]$ in  \eqref{eq:CanChargesVariation} by setting the corresponding gauge parameter to one and the other one to zero. Or equivalently one can write this in a little bit more suggestive form as
    \begin{equation}\label{eq:ElectricChargeVariation}
        \delta Q_\mathcal{J} :=\frac{\kappa_P}{8\pi}\int\extd\varphi\langle \Jt\delta\mathcal{C}_\varphi\rangle,\qquad\delta Q_\mathcal{P} :=\frac{\kappa_P}{8\pi}\int\extd\varphi\langle \Pt\delta\mathcal{C}_\varphi\rangle.
    \end{equation}
Similar to mass and angular momentum the electric charges are given by
    \begin{equation}
        \delta Q_\mathcal{J} = \delta\Jt_0,\qquad\delta Q_\mathcal{P} = \delta\Pt_0.
    \end{equation}
After having determined mass, angular momentum and electric charges the remaining pieces of the puzzle to calculate the entropy via the first law \eqref{eq:FirstLaw} is to determine the thermodynamic potentials. We will do so by imposing conditions on the holonomy $e^{i(h+\bar{h})}$ with
    \begin{subequations}\label{eq:HolonomyDefinition}
    \begin{align}
        h & =-\frac{\beta_T}{2\pi}\left(\int\extd\varphi\,a_u - \Omega \int\extd\varphi\,a_\varphi\right),\label{eq:HolCondA}\\
        \bar{h} & =-\frac{\beta_T}{2\pi}\left(\int\extd\varphi\,c_u - \Omega \int\extd\varphi\,c_\varphi+\Phi_\mathcal{J} \int\extd\varphi\,\Jt+\Phi_\mathcal{P} \int\extd\varphi\,\Pt\right)\label{eq:HolCondB}.
    \end{align}
    \end{subequations}
Before stating what these conditions are it may be illuminating to rewrite the first law \eqref{eq:FirstLaw} using \eqref{eq:VariationMassAngMomFormula} and \eqref{eq:ElectricChargeVariation} as well as \eqref{eq:HolonomyDefinition} to yield
    \begin{equation}\label{eq:EntropyAndHol}
        \delta S_{\textrm{Th}} = k\langle h\,\delta a_\varphi\rangle+\frac{\kappa_P}{2}\langle \bar{h}\,\delta c_\varphi\rangle.
    \end{equation}
We want to point out that aside from assuming that the solution in question satisfies a first law we did not make any additional assumptions.\\
Looking at \eqref{eq:HolCondA} one sees that $h$ is noting else than the holonomies of (uncharged) rotating cosmological solutions in flat space (see e.g. \cite{Detournay:2016sfv,Ammon:2017vwt} or for the AdS case \cite{deBoer:2013gz}). Thus demanding that the holonomy associated to $h$ lies in the center of the gauge group\footnote{To be more precise, one has $b^{-1}e^{ih}b=-\unity$} is tantamount to demanding that the eigenvalues of $h$ satisfy $\textrm{Eigen}\left[h\right]= \textrm{Eigen}\left[2\pi \Lt_0\right]$. Now looking at \eqref{eq:HolCondB} it is natural to demand that the holonomy of $\bar{h}$ lies in the center of the gauge group as well. This requirement can also be motivated from higher-spin theories both in AdS, as well as in flat space. If there are additional higher-spin charges present one demands that the total holonomies including all charges has to coincide with the holonomies of the uncharged solutions (see e.g. \cite{Gutperle:2011kf,Ammon:2011nk}). Thus, in order for the holonomies $h$ and $\bar{h}$ to lie in the center of the gauge group they have to satisfy
    \begin{equation}\label{eq:HolCond}
        \textrm{Eigen}\left[h\right]= \textrm{Eigen}\left[2\pi \Lt_0\right],\qquad\textrm{and}\qquad\textrm{Eigen}\left[\bar{h}\right]= 0.
    \end{equation}
As a consequence this also means that for the case where \eqref{eq:HolCond} is satisfied one can also functionally integrate \eqref{eq:EntropyAndHol} to yield
    \begin{equation}\label{eq:EntropyAndHol2}
        S_{\textrm{Th}} = k\langle h\,a_\varphi\rangle+\frac{\kappa_P}{2}\langle\bar{h}\,c_\varphi\rangle.
    \end{equation}  
Imposing the conditions \eqref{eq:HolCond} yields the following relations for the inverse temperature $\beta_T$ and angular velocity $\Omega$ after integrating over $\varphi$ in \eqref{eq:HolonomyDefinition}
    \begin{equation}\label{eq:InvTempAndAngVel}
        \beta_T=\frac{\pi\left(\Lt_0-2\frac{\Jt_0\Pt_0}{\kappa_P}\right)}{\left(\Mt_0-\frac{\Pt_0^2}{\kappa_P}\right)\sqrt{\frac{2}{k}\left(\Mt_0-\frac{\Pt_0^2}{\kappa_P}\right)}},\qquad \Omega=\frac{2\left(\Mt_0-\frac{\Pt_0^2}{\kappa_P}\right)}{\Lt_0-2\frac{\Jt_0\Pt_0}{\kappa_P}}.
    \end{equation}
The electric potentials are given by  
    \begin{equation}\label{eq:ElectricPotentials}
        \Phi_\mathcal{J}=\frac{4\Pt_0\left(\Mt_0-\frac{\Pt_0^2}{\kappa_P}\right)}{\kappa_P\left(\Lt_0-2\frac{\Jt_0\Pt_0}{\kappa_P}\right)},\qquad \Phi_\mathcal{P}=\frac{2\left(2\Mt_0\Jt_0-\Lt_0\Pt_0\right)}{\kappa_P\left(\Lt_0-2\frac{\Jt_0\Pt_0}{\kappa_P}\right)}.
    \end{equation}
Now using \eqref{eq:EntropyAndHol2} one immediately finds the thermal entropy to be
    \begin{equation}\label{eq:FlowInvEntropy}
        S_{\textrm{Th}}=\frac{\pi}{6}\frac{c_M\left(\Lt_0-2\frac{\Jt_0\Pt_0}{\kappa_P}\right)}{\sqrt{\frac{c_M}{6}\left(\Mt_0-\frac{\Pt_0^2}{\kappa_P}\right)}},
    \end{equation}
which is exactly the same result as in \eqref{eq:ThermalEntropyArea}.This shows that the holonomy conditions \eqref{eq:HolCond} were, indeed, a sensible choice to make in the Chern-Simons formulation.\\
For the SUSY example described in Section~\ref{sec:N4FlatSupergravity} one can use again the first law \eqref{eq:FirstLaw} in order to determine the thermal entropy of flat space cosmologies in a SUSY background. As such one can also use the same kind of reasoning as before with some minor modifications. These modifications mainly concern the (variation of) mass, angular momentum and the electric charges. For the $\mathcal{N}=4$ SUSY case the variations of the charges characterizing the cosmological solution are given by
    \begin{subequations}\label{eq:VariationMassAngMomFormulaSUSY}
    \begin{align}
        \delta M &:= \delta Q[\epsilon|_{\partial_u}]+\delta Q[\bar{\epsilon}|_{\partial_u}]=\frac{k}{2\pi}\int\extd\varphi\langle\mathcal{A}_u\delta\mathcal{A}_\varphi\rangle=\delta\Mt_0,\\
        \delta J &:= \delta Q[\epsilon|_{\partial_\varphi}]+\delta Q[\bar{\epsilon}|_{\partial_\varphi}]=\frac{k}{2\pi}\int\extd\varphi\langle\mathcal{A}_\varphi\delta\mathcal{A}_\varphi\rangle=\delta\Lt_0,\\
        \delta Q_\mathcal{J} & :=\frac{k}{2\pi}\int\extd\varphi\langle \Jt\delta\mathcal{A}_\varphi\rangle=\delta\Jt_0,\\
        \delta Q_\mathcal{P} &=\frac{k}{2\pi}\int\extd\varphi\langle \Pt\delta\mathcal{A}_\varphi\rangle=\delta\Pt_0.
    \end{align}
    \end{subequations}
One can now again impose that the holonomy
    \begin{equation}
        h =-\frac{\beta_T}{2\pi}\left(\int\extd\varphi\,a_u - \Omega \int\extd\varphi\,a_\varphi+\Phi_\mathcal{J} \int\extd\varphi\,\Jt+\Phi_\mathcal{P} \int\extd\varphi\,\Pt\right),
    \end{equation}
satisfies
    \begin{equation}\label{eq:HolCondSUSY}
        \textrm{Eigen}\left[h\right]= \textrm{Eigen}\left[2\pi \Lt_0\right].
    \end{equation}
This again allows to fix the inverse temperature $\beta_T$, angular potential $\Omega$ and electrical potentials $\Phi_\mathcal{J}$ and $\Phi_\mathcal{P}$ that again exactly correspond to \eqref{eq:InvTempAndAngVel} and \eqref{eq:ElectricPotentials} but with $\kappa_P=8k$. Consequently also the thermal entropy is given by \eqref{eq:FlowInvEntropy} with $\kappa_P=8k$.\\
For the ``reloaded'' cases described in Section~\ref{sec:Reloaded} one can apply the exact same steps that have been employed previously in this section in order to obtain the thermal entropy of cosmological solutions in these theories. The only thing one has to take care of is that the angular momentum $J$ as well as the $\hat{\mathfrak{u}}(1)$ charge $Q_\mathcal{J}$ are modified due to the different bilinear form and thus one has to replace $\Lt_0$ and $\Jt_0$ in \eqref{eq:FlowInvEntropy} with the zero modes of the tilded expressions defined in \eqref{eq:ReloadedEinsteinCharges} and \eqref{eq:ReloadedSUSYCharges}. By doing so one obtains
    \begin{equation}\label{eq:ReloadedThermalEntropy}
        \boxed{S_{\textrm{Th}}=\frac{\pi}{6}\frac{c_M\left(\Lt_0-2\frac{\Jt_0\Pt_0}{\kappa_P}+\frac{\kappa_J \Pt_0^2}{\kappa_P^2}\right)+c_L\left(\Mt_0-\frac{\Pt_0^2}{\kappa_P}\right)}{\sqrt{\frac{c_M}{6}\left(\Mt_0-\frac{\Pt_0^2}{\kappa_P}\right)}},}
    \end{equation}
where the central charges as well as the $\hat{\mathfrak{u}}(1)$ levels take the same values as described in Section~\ref{sec:Reloaded} depending on the specific model in question. This expression for the entropy of cosmological solutions in the presence of $\hat{\mathfrak{u}}(1)$ charges is one of the main results of this work.

%--------------------------------------------------
\section{Thermal Entropy: Field Theory Side} \label{sec:ThermalField}
%--------------------------------------------------

This section is focused on determining the thermal entropy of quantum field theories with the general symmetries given by \eqref{eq:BMSU1Squared} and \eqref{n4subms} via a corresponding partition function. We will present two ways of computing the partition function in this section. One makes use of the powerful spectral flow automorphism that we found previously whereas the other one uses a saddlepoint approximation. In the Einstein gravity case one can apply in principle both techniques, however, as we will see using the spectral flow automorphism turns out to be the more efficient method for this case. In the $\mathcal{N}=4$ flat supergravity case, however, the saddlepoint approximation is the method of choice. In the same spirit we determine the logarithmic corrections to the thermal entropy as well. 

%--------------------------------------------------
\subsection{Partition Functions and Spectral Flow}
%--------------------------------------------------

Let us consider the following ensemble:
    \begin{equation}\label{eq:BMSEnsemble}
        Z(\rho,\eta,\mu,\nu)=\textnormal{Tr}\,e^{2\pi i \left(\rho \Mt_0+\eta \Lt_0+\mu \Pt_0+\nu \Jt_0\right)},
    \end{equation}
where $2\pi i\rho=-\beta_T$ and $2\pi\eta=\Omega$ encode the inverse temperature and the angular potential respectively and $\mu$ and $\eta$ are the chemical potentials associated to the $\hat{\mathfrak{u}}(1)$ charges.\\
Furthermore, we assume that the trace in \eqref{eq:BMSEnsemble} is taken in a highest-weight \eqref{eq:HilbertEigenvalues} representation\footnote{In most of the literature on flat space holography in 3D so far highest weight representations have been quite successfully used for checks of a holographic duality involving asymptotically flat spacetimes such as in e.g. \cite{Bagchi:2014iea,Bagchi:2015wna,Basu:2015evh,Bagchi:2012xr}. However, the "caveat" of these representations is that they are non-unitary as soon as $c_M\neq0$ \cite{Bagchi:2009pe}. Alternatively \emph{induced} representations that do not suffer this caveat have been proposed as more suitable representations on the quantum field theory side (including possible higher-spin extensions) of the proposed holographic duality \cite{Barnich:2014kra,Campoleoni:2015qrh,Campoleoni:2016vsh}. Even though these two representations are quite different there is one particular thing that they have in common and that is that in both representations states can be labelled as in \eqref{eq:HilbertEigenvalues}.}.\\  
We now derive a Cardy-like formula for charged flat space cosmologies in 3D flat space by making use of the spectral flow \eqref{eq:BMSFlow} as well as $\mathfrak{bms}_3$ modular transformations. As a first step we apply the (inverse) flow \eqref{eq:BMSFlow} to \eqref{eq:BMSEnsemble}. For the following choice of spectral flow parameters:
    \begin{equation}
        u=\frac{\nu}{\eta},\qquad v=\frac{\eta\mu-\nu\rho}{\nu^2},
    \end{equation}   
the partition function \eqref{eq:BMSEnsemble} becomes
    \begin{equation}\label{eq:PartFunctionAfterFlow}
         Z(\rho,\eta,\mu,\nu)=\exp\left[-i\pi\left(\frac{\nu^2}{2\eta}\kappa_J+\frac{2\eta\mu \nu-\nu^2\rho}{2\eta^2}\nu\kappa_P\right)\right]\textrm{Tr}\,e^{2\pi i\left(\rho \tilde{\Mt}_0+\eta \tilde{\Lt}_0\right)}.
    \end{equation}
Since the spectral flow is an automorphism of the algebra the right-hand side of \eqref{eq:PartFunctionAfterFlow} is equivalent to the original partition function \eqref{eq:BMSEnsemble}. The only thing left to compute is the remaining trace in \eqref{eq:PartFunctionAfterFlow}. Thus instead of using invariance of the partition function under the $\mathfrak{bms}$ equivalent of modular transformations (see e.g. \eqref{eq:ModSTrafo}) and then using a saddlepoint approximation to determine the leading order contribution to the partition function the presence of the two-parameter spectral flow allows for a much more efficient treatment of the problem at hand. Determining the remaining trace in \eqref{eq:PartFunctionAfterFlow} is then a straightforward process that we will outline in the following \cite{Bagchi:2012xr}.\\
One can now use the invariance of $\textrm{Tr}\,e^{2\pi i\left(\rho \tilde{\Mt}_0+\eta \tilde{\Lt}_0\right)}$ under $\mathfrak{bms}_3$ modular transformations \cite{Bagchi:2012xr}
    \begin{equation}
        \rho\rightarrow\frac{\rho}{\eta^2},\qquad\eta\rightarrow-\frac{1}{\eta},
    \end{equation}
in order to obtain
    \begin{equation}
         Z(\rho,\eta,\mu,\nu)=\exp\left[-i\pi\left(\frac{\nu^2}{2\eta}\kappa_J+\frac{2\eta\mu \nu-\nu^2\rho}{2\eta^2}\nu\kappa_P\right)\right]\textrm{Tr}\,e^{\frac{2\pi i}{\eta}\left(\frac{\rho}{\eta} \tilde{\Mt}_0-\tilde{\Lt}_0\right)}.
    \end{equation}
%    \begin{equation}
%        \log Z(\rho,\eta,\mu,\nu)=2\pi i\left(\frac{\rho}{\eta^2}\tilde{M}_0^{\textrm{min}}-\frac{\tilde{L}_0^{\textrm{min}}}{\eta}-\frac{\nu^2}{4\eta}\kappa_J-\frac{2\eta\mu \nu-\nu^2\rho}{4\eta^2}\nu\kappa_P\right).
%    \end{equation}
For small values of $\frac{2\pi i\rho}{\eta^2}=-4\pi^2\frac{\beta_T}{\Omega^2}$, i.e. for high temperatures at some given value of the angular potential only the vacuum with energy $\tilde{\Mt}_0^{\textrm{min}}=h_M^v$ and corresponding $\tilde{\Lt}_0^{\textrm{min}}=h_L^v$ contributes and so one obtains
    \begin{equation}\label{eq:MagicLogZ}
        \log Z(\rho,\eta,\mu,\nu)=\frac{2\pi i}{\eta}\left(\frac{\rho}{\eta}h_M^v-h_L^v-\frac{\nu^2}{4}\kappa_J-\frac{2\eta\mu \nu-\nu^2\rho}{4\eta}\kappa_P\right).
    \end{equation}
In the microcanonical ensemble one has
    \begin{subequations}\label{eq:ChargesAndPotentials}
    \begin{align}
        h_L & = \frac{1}{2\pi i}\partial_\eta\log Z = \frac{h_L^v}{\eta^2}-2\frac{\rho h_M^v}{\eta^3}+\frac{\nu^2\kappa_J}{4\eta^2}+\frac{\nu(\eta\mu-\nu\rho)\kappa_P}{2\eta^3},\\
        h_M & = \frac{1}{2\pi i}\partial_\rho\log Z = \frac{h_M^v}{\eta^2}+\frac{\nu^2\kappa_P}{4\eta^2},\\
        j & = \frac{1}{2\pi i}\partial_\nu\log Z = -\frac{\nu\kappa_J}{2\eta}-\frac{(\eta\mu-\nu\rho)\kappa_P}{2\eta^2},\\
        p & = \frac{1}{2\pi i}\partial_\mu\log Z = -\frac{\nu\kappa_P}{2\eta},
    \end{align}
    \end{subequations}
and therefore also
    \begin{equation}\label{eq:LegendreTrafoEntropy}
        S_{\textrm{Th}}=(-1+\eta\partial_\eta+\rho\partial_\rho+\nu\partial_\nu+\mu\partial_\mu)\log Z = 4\pi i\frac{\eta h_L^v-\rho h_M^v}{\eta^2},
    \end{equation}
or in terms of $h_L$, $h_M$, $j$ and $p$ and replacing $h_L^v=-\frac{c_L}{24}$, $h_M^v=-\frac{c_M}{24}$
    \begin{equation}\label{eq:FieldTheoryEntropy}
        \boxed{S_{\textrm{Th}}=\frac{\pi}{6}\frac{c_M\left(h_L-2\frac{jp}{\kappa_P}+\frac{\kappa_J p^2}{\kappa_P^2}\right)+c_L\left(h_M-\frac{p^2}{\kappa_P}\right)}{\sqrt{\frac{c_M}{6}\left(h_M-\frac{p^2}{\kappa_P}\right)}},}
    \end{equation}
which is exactly the same expression as \eqref{eq:ReloadedThermalEntropy}.\\    
Since this expression contains terms that are inverse proportional to $\kappa_P$ a natural question to ask is how this expression for the entropy looks like for $\kappa_P=0$. Since \eqref{eq:LegendreTrafoEntropy} does not contain $\kappa_P$ the relevant equations for $\kappa_P=0$ are \eqref{eq:ChargesAndPotentials}. For this case these equations read
    \begin{subequations}\label{eq:ChargesAndPotentialsKappaP0}
    \begin{align}
        h_L & = \frac{1}{2\pi i}\partial_\eta\log Z = \frac{h_L^v}{\eta^2}-2\frac{\rho h_M^v}{\eta^3}+\frac{\nu^2\kappa_J}{4\eta^2},\\
        h_M & = \frac{1}{2\pi i}\partial_\rho\log Z = \frac{h_M^v}{\eta^2},\\
        j & = \frac{1}{2\pi i}\partial_\nu\log Z = -\frac{\nu\kappa_J}{2\eta},\\
        p & = 0.
    \end{align}
    \end{subequations}
Thus solving these equations for in terms of $\eta$, $\rho$, $\nu$ and inserting the result into \eqref{eq:LegendreTrafoEntropy} one obtains
    \begin{equation}
        S_{\textrm{Th}}=\frac{\pi}{6}\frac{c_M\left(h_L-\frac{j^2}{\kappa_J}\right)+c_Lh_M}{\sqrt{\frac{c_M}{6}h_M}}.
    \end{equation}
It is also noteworthy that this expression is invariant under the flow \eqref{eq:BMSFlow} with $\kappa_P=0$ and $v=0$.\\
Since \eqref{eq:LegendreTrafoEntropy} does not depend on $\kappa_P$ and $\kappa_J$ one can use the same logic as before, i.e. set $\kappa_J=0$ in \eqref{eq:ChargesAndPotentialsKappaP0} and recover the thermal entropy of a $\mathfrak{bms}_3$ invariant field theory without any additional $\hat{\mathfrak{u}}(1)$ currents.\\
Having determined the thermal entropy for a quantum field theory with underlying extended $\mathfrak{bms}_3$ symmetry given by \eqref{eq:BMSU1Squared} we now focus on a quantum field theory with $\mathcal{N}=4$ super $\mathfrak{bms}_3$ symmetry \eqref{n4subms}. In close analogy to general supersymmetric quantum field theories we propose the following elliptic genus \cite{Schellekens:1986xh,Witten:1986bf} as a partition function for $\mathcal{N}=4$ super $\mathfrak{bms}_3$ invariant quantum field theories:
    \begin{equation}\label{eq:PartitionFunctionSUSY}
        Z(\rho, \eta, \mu, \nu) = \mathrm{Tr}_{RR}(-1)^F e^{2 \pi i \left( \rho \Mt_0 + \eta \Lt_0 + \mu \Pt_0 + \nu \Jt_0\right)},
    \end{equation}
where $F$ is a fermionic number operator, $\mathrm{Tr}_{RR}$ means that the trace is taken in the Ramond-Ramond sector of the quantum field theory and the parameters $\rho$, $\eta$, $\mu$ and $\nu$ are related to inverse temperature, angular potential and chemical potentials as in the purely bosonic case treated previously.\\
Since the $\mathcal{N}=4$ super $\mathfrak{bms}_3$ algebra exhibits spectral flow symmetry any physical observable should also be invariant under spectral flow. Using this invariance one can show that the following relation holds:
    \begin{equation}
        Z(\rho, \eta, \mu, \nu) = \exp\left[\frac{\pi i}{2} \big( \left(\rho + 2 \mu \right)\kappa_P + \left(\eta + 2 \nu \right)\kappa_J\big)\right] Z(\rho, \eta, \mu+ \rho, \nu + \eta),
    \end{equation}
since both sides of the equation can be related via a spectral flow \eqref{reducedflow} with $u=-1$. It should be noted that the flow \eqref{reducedflow} is  a sub-set of the two-parameter flow encountered earlier in \eqref{eq:BMSFlow}. This can be seen by only taking the $u$-flow in \eqref{eq:BMSFlow} while freezing the $v$-flow and setting $ \kappa_P = \tfrac{2 c_{M}}{3}$ and $\kappa_J =\tfrac{2 c_{L}}{3}$. This is natural, as \eqref{eq:BMSU1Squared} is just the bosonic sub-algebra of \eqref{n4subms} with these identifications.\\
Since in the present case we only have the reduced spectral flow \eqref{reducedflow} and not the full general flow \eqref{eq:BMSFlow} at our disposal one cannot quite use the same convenient techniques used in the previous case. This is, however, not a problem since one can still apply other techniques to get an explicit expression for the partition function \eqref{eq:PartitionFunctionSUSY}. In more physical terms, the partition function \eqref{eq:PartitionFunctionSUSY} is a sum over all microstates in a given theory weighted with an appropriate weight factor that is given by the exponential factor in \eqref{eq:PartitionFunctionSUSY}. Thus at high temperatures the partition function can be related to the density of states $d(h_L,h_M,j,p)$ by
    \begin{equation}\label{eq:DOSdef}
       Z(\rho,\eta,\mu,\nu)=\int\extd h_L\extd h_M\extd j\extd p\,d(h_L,h_M,j,p)e^{2\pi i(\rho h_M+\eta h_L+\mu p+\nu j)},
    \end{equation}
where $h_L,\,h_M,\,j$ and $p$ are the eigenvalues of $\Lt_0,\,\Mt_0,\,\Jt_0$ and $\Pt_0$ respectively. Invariance of the partition function under the flat space analogue of modular transformations requires the partition function to satisfy\footnote{One way to see this is to take a suitable limit along the lines of \cite{Bagchi:2012xr} of the transformation behaviour of an $\mathcal{N}=4$ superconformal partition function as described in \cite{Kawai:1993jk}.}
    \begin{equation}\label{eq:ModSTrafo}
        Z(\rho,\eta,\mu,\nu)=e^{-i\pi\left(\frac{\nu^2}{2\eta}\kappa_J+\frac{2\eta\mu \nu-\nu^2\rho}{2\eta^2}\kappa_P\right)}Z\left(\frac{\rho}{\eta^2},-\frac{1}{\eta},\frac{\eta\mu-\nu\rho}{\eta^2},\frac{\nu}{\eta}\right).
    \end{equation}
This implies that the vacuum state dominates the partition function at high temperatures
    \begin{equation}
        Z(\rho,\eta,\mu,\nu)\approx e^{-i\pi\left(\frac{\nu^2}{2\eta}\kappa_J+\frac{2\eta\mu\nu -\nu^2 \rho}{2\eta^2}\nu\kappa_P\right)}e^{2\pi i\left(\frac{\rho}{\eta^2}h_M^v-\frac{1}{\eta}h_L^v\right)},
    \end{equation}
where we have assumed that the vacuum state is electrically neutral $j^v=p^v=0$. Inverting \eqref{eq:DOS} one obtains
    \begin{equation}\label{eq:DOS}
       d(h_L,h_M,j,p)=\int\extd \rho\extd \eta\extd \mu\extd \nu e^{-i\pi\left(\frac{\nu^2}{2\eta}\kappa_J+\frac{2\eta\mu\nu-\nu^2\rho}{2\eta^2}\kappa_P\right)}e^{2\pi i\left(\frac{\rho}{\eta^2}h_M^v-\frac{1}{\eta}h_L^v-\rho h_M-\eta h_L-\mu p-\nu j\right)}.
    \end{equation}
This integral can then be evaluated by saddle point methods. This integrand reaches an extremum at
    \begin{subequations}\label{eq:Saddle}
    \begin{align}
        \rho_0 & =i\left( \frac{h_L^v}{2\sqrt{h_M^v\left(\frac{p^2}{\kappa_P}-h_M\right)}}+\frac{h_M^v\left(p^2\frac{\kappa_J}{\kappa_P}-2jp+h_L\kappa_P\right)}{2\kappa_P\sqrt{h_M^v\left(\frac{p^2}{\kappa_P}-h_M\right)^3}}\right),\\
        \eta_0 & =i\sqrt{\frac{h_M^v}{\frac{p^2}{\kappa_P}-h_M}},\\
        \mu_0 & =-i\left(\frac{h_L^vp}{\kappa_P}\frac{1}{\sqrt{h_M^v\left(\frac{p^2}{\kappa_P}-h_M\right)}}-\frac{h_M^v\left(p^3\frac{\kappa_J}{\kappa_P}+2h_Mj\kappa_P-p\left(2h_M\kappa_J+h_L\kappa_P\right)\right)}{\kappa_P^2\sqrt{h_M^v\left(\frac{p^2}{\kappa_P}-h_M\right)^3}}\right),\\
        \nu_0 & =-i\frac{2p}{\kappa_P}\sqrt{\frac{h_M^v}{\frac{p^2}{\kappa_P}-h_M}}.
    \end{align}
    \end{subequations}
Using the saddle \eqref{eq:Saddle} one obtains for the entropy
    \begin{equation}
        S_{\textrm{Th}}=\log d_{(0)} = -\pi\frac{4h_M^v\left(h_L-2\frac{jp}{\kappa_P}+\frac{\kappa_J p^2}{\kappa_P^2}\right)+4h_L^v\left(h_M-\frac{p^2}{\kappa_P}\right)}{\sqrt{-4h_M^v\left(h_M-\frac{p^2}{\kappa_P}\right)}}.
    \end{equation}
Setting
    \begin{equation}
        h_L^v=-\frac{c_L}{24},\qquad h_M^v=-\frac{c_M}{24},
    \end{equation}
this expression exactly coincides with \eqref{eq:FieldTheoryEntropy}.

%--------------------------------------------------
\subsection{Logarithmic Corrections}
%--------------------------------------------------

Since the previous subsection (without SUSY) did not explicitly rely on a saddle point approximation it would be nice if we could also get the logarithmic corrections to the entropy in a similar manner as before. Thus it might be sensible to look at a similar quantity as in \eqref{eq:SaddleLog}. In our case it is very suggestive to look at the determinant of the matrix
    \begin{equation}
        \mathcal{Z}_{ab}=\frac{1}{2\pi i}\partial_a\partial_b\log Z,
    \end{equation}
where the derivatives are taken with respect to the potentials $\rho$, $\eta$, $\mu$ and $\nu$. This matrix may also be written in a little bit more suggestive form as
    \begin{equation}
        \mathcal{Z}_{ab}=
        \left(
			\begin{array}{cccc}
				 \partial_\rho h_M & \partial_\eta h_M & \partial_\mu h_M & \partial_\nu h_M\\
				\partial_\rho h_L & \partial_\eta h_L & \partial_\mu h_L & \partial_\nu h_L\\
				\partial_\rho p & \partial_\eta p & \partial_\mu p & \partial_\nu p\\
				\partial_\rho j & \partial_\eta j & \partial_\mu j & \partial_\nu j
			\end{array}\right).
    \end{equation}
Then the logarithmic correction to the thermal entropy is given by
    \begin{equation}\label{eq:MagicLogCorr}
        \Delta S_{\textrm{Th}}=-\frac{1}{2}\log\det\mathcal{Z}.
    \end{equation}
This can be motivated by writing \eqref{eq:DOS} as
    \begin{equation}
       d(h_L,h_M,j,p)=\int\extd \rho\extd \eta\extd \mu\extd \nu e^{2\pi if(\rho,\eta,\mu,\nu)},
    \end{equation}
with
    \begin{equation}\label{eq:MagicConnection}
        f(\rho,\eta,\mu,\nu)=\frac{1}{2\pi i}\log Z-(\rho h_M+\eta h_L+\mu p+\nu j).
    \end{equation}
The logarithmic corrections to the thermal entropy can then be determined by looking at
    \begin{equation}
        \Delta S_{\textrm{Th}}=-\frac{1}{2}\log\det\partial_a\partial_b f,
    \end{equation}
where this expression is evaluated at the saddle point. However, looking at \eqref{eq:MagicConnection} one can immediately see that
    \begin{equation}
        \partial_a\partial_bf=\frac{1}{2\pi i}\partial_a\partial_b\log Z.
    \end{equation}
Furthermore solving \eqref{eq:ChargesAndPotentials} in terms of the potentials $\rho$, $\eta$, $\mu$ and $\nu$ one recovers precisely the relations of the saddle \eqref{eq:Saddle}. This shows that our assumption is, indeed, valid.\\  
Now applying \eqref{eq:MagicLogCorr} for \eqref{eq:MagicLogZ} and using \eqref{eq:ChargesAndPotentials} one obtains
    \begin{equation}\label{eq:LogCorrection}
        \Delta S_{\textrm{Th}} = -\frac{1}{2}\log\left[\frac{\left(h_M^v\right)^2\kappa_P^2}{\eta^8}\right]=-\log\left[\frac{24\kappa_P\left(\frac{p^2}{\kappa_P}-h_M\right)^2}{c_M}\right].
    \end{equation}
As expected, this expression only contains flow invariant quantities and is valid for both Einstein gravity and the $\mathcal{N}=4$ supergravity case as well as their ``reloaded'' versions. It is also worthwhile noting that for Einstein gravity, the $\mathcal{N}=4$ supergravity case as well as their ``reloaded'' versions the thermal entropy including the logarithmic corrections \eqref{eq:LogCorrection} can be written as
    \begin{equation}\label{eq:CorrectedEntropyCorefficient}
        S_{\textrm{Th}}=S_{(0)}-q \log\left[S_{(0)}\right]+q\log\left[\beta_T\right]+\ldots,
    \end{equation}
where $S_{(0)}$ is the leading order piece of the entropy, $\beta_T$ the inverse temperature and  the coefficient in front of the logarithm is $q=2$.\\    
In case of vanishing $\kappa_P$ one can perform the same calculation, however, taking into account that $p$ as well as $\mu$ are now zero. This means in particular that $\mathcal{Z}_{ab}$ is now a $3\times3$ matrix of the form       
    \begin{equation}
        \mathcal{Z}_{ab}=
        \left(
			\begin{array}{ccc}
				 \partial_\rho h_M & \partial_\eta h_M &  \partial_\nu h_M\\
				\partial_\rho h_L & \partial_\eta h_L &  \partial_\nu h_L\\
				\partial_\rho j & \partial_\eta j &  \partial_\nu j
			\end{array}\right).
    \end{equation} 
For this case and again using \eqref{eq:ChargesAndPotentials} with $\kappa_P=0$ one obtains    
    \begin{equation}
        \Delta S_{\textrm{Th}} = -\frac{1}{2}\log\left[\frac{2\left(h_M^v\right)^2\kappa_J}{\eta^7}\right]=-\frac{1}{4}\log\left[4^4\kappa_J^2h_M^7\left(\frac{6}{c_M}\right)^3\right].
    \end{equation}
This expression can again be expressed in the general form \eqref{eq:CorrectedEntropyCorefficient} using the appropriate expressions for the leading piece of the entropy $S_{(0)}$ as well as the inverse temperature $\beta_T$ where now one has $q=\frac{7}{4}$.\\
A very good cross check as to whether or not the way we compute things make sense is to see what happens if we also set $\kappa_J$ to zero. In that case we should recover precisely the results of \cite{Bagchi:2013qva} where the logarithmic corrections to the thermal entropy of flat space cosmologies have been determined.\\
Now for $\kappa_J=\kappa_P=0$ the matrix $\mathcal{Z}$ takes the form
    \begin{equation}
        \mathcal{Z}_{ab}=
        \left(
			\begin{array}{cc}
				 \partial_\rho h_M & \partial_\eta h_M\\
				\partial_\rho h_L & \partial_\eta h_L
			\end{array}\right).
    \end{equation}
Using again \eqref{eq:ChargesAndPotentials} but now with $\kappa_J=\kappa_P=0$ we obtain
    \begin{equation}
        \Delta S_{\textrm{Th}} = -\frac{1}{2}\log\left[-\frac{4\left(h_M^v\right)^2}{\eta^6}\right]=-\frac{3}{2}\log\left[2h_M\sqrt{\frac{12}{c_M}}\right],
    \end{equation}
which is exactly the logarithmic correction\footnote{There is a factor 12 difference with respect to the results in \cite{Bagchi:2013qva}. This, however, is only due to a slightly different definition of $c_M$. Replacing $c_M\rightarrow12c_M$ one immediately recovers (4.14) in \cite{Bagchi:2013qva}.} of the entropy of a flat space cosmology found in \cite{Bagchi:2013qva}.

%--------------------------------------------------
\section{Conclusions and Outlook}
%--------------------------------------------------

In this paper we explored spectral flow symmetry in the context of flat space holography. We showed spectral flow invariance of certain $\hat{\mathfrak{u}}(1)$ extended (supersymmetric) $\mathfrak{bms}_3$ algebras and presented specific models of gravity along with suitable boundary conditions whose asymptotic symmetries are given by said extended $\mathfrak{bms}_3$ algebras. We then determined the thermal entropy of cosmological solutions in these models using both the Bekenstein-Hawking area law as well as integrating the first law of flat space cosmologies. In addition we derived the thermal entropy (including logarithmic corrections) from a putative dual quantum field theory partition function.\\
There are a couple things that would be interesting to do following up this work. One thing might be to look for more gravity models exhibiting spectral flow symmetry in other contexts such as e.g. TMG or flat space chiral gravity \cite{Bagchi:2012yk}. Of course one does not necessarily have to be restricted to just $\hat{\mathfrak{u}}(1)$ extensions of (super) $\mathfrak{bms}_3$. For example $\hat{\mathfrak{su}}(N)$ extensions should exhibit spectral flow as well and it would be very interesting to work this out explicitly.\\
Furthermore it would be good to better understand why there is only a one-parameter spectral flow in the $\mathcal{N}=4$ super $\mathfrak{bms}_3$ algebra present as in comparison to the purely bosonic subalgebra. From a naive point of view one would have expected a two-parameter flow both from the perspective of a \.In\"on\"u--Wigner contraction of the $\mathcal{N}=(2,2)$ super Virasoro algebra as well as from the $\mathfrak{bms}_3\inplus\hat{\mathfrak{u}}(1)\inplus\hat{\mathfrak{u}}(1)$ case. However, it is not obvious at all how the second parameter spectral flow could emerge in the $\mathcal{N}=4$ super $\mathfrak{bms}_3$ case. This is something that still needs to be better understood.\\
Recently there has also been progress in gaining a better understanding of a holographic principle in asymptotically flat spacetimes in 3D on the full asymptotic boundary of flat space by linking future and past null infinity\footnote{For four and higher, even dimensions, this has been first worked out in \cite{Strominger:2013jfa,Kapec:2015vwa}, with some objections argued in \cite{Hollands:2016oma}.} \cite{Prohazka:2017equ,Compere:2017knf}. It would be interesting to see if similar arguments could also be used for the cases treated in this work that include additional gauge fields.

%---------------------------------------------------------------------------------------------------------------
\subsection*{Acknowledgments}
%---------------------------------------------------------------------------------------------------------------

We are grateful to Arjun Bagchi, Alejandra Castro, Oscar Fuentealba, Mirah Gary, Daniel Grumiller, Diego Hofman, Wout Merbis, Jan Rosseel and Ricardo Troncoso for enlightening discussions and comments. The work of RB is supported by Inspire Faculty scheme, DST India and in part by the Belgian Federal Science Policy Office (BELSPO) through the Interuniversity Attraction Pole P7/37 and a research-and-return grant, in part by the ``FWO-Vlaanderen" through the project G020714N, and by the Vrije Universiteit Brussel through the Strategic Research Program ``High-Energy Physics". SD and MR thank the Galileo Galilei Institute for Theoretical Physics, and MR thanks the Abdus Salam International Centre for Theoretical Physics for the hospitality and the Istituto Nazionale di Fisica Nucleare (INFN) as well as the Italian Banking Foundation
Association for partial support during the completion of this work. Furthermore MR is grateful for the hospitality of the Erwin Schr{\"o}dinger International Institute (ESI) for Mathematics and Physics. SD is a Research Associate of the Fonds de la Recherche Scientifique F.R.S.-FNRS (Belgium). He is supported in part by the ARC grant ``Holography, Gauge Theories and Quantum Gravity Building models of quantum black holes'', by IISN - Belgium (convention 4.4503.15) and benefited from the support of the Solvay Family. The research of MR is supported by the ERC Starting Grant 335146 "HoloBHC".\\\\

%%%%%%%%%%%%%%%%%%%%%%%%%%%%%%%%%%%%%%%%%%%%%%%%%%%%%%%%%%%%%%%%%%%%%
%%%%%%%%%%%%%%%%%%%%%%%%%%%%% APPENDIX %%%%%%%%%%%%%%%%%%%%%%%%%%%%
%%%%%%%%%%%%%%%%%%%%%%%%%%%%%%%%%%%%%%%%%%%%%%%%%%%%%%%%%%%%%%%%%%%%%

\begin{appendix}

%--------------------------------------------------
\section{Matrix Representations}\label{sec:MatrixRepresentations}
%--------------------------------------------------

In this section we collect some explicit matrix representations of the algebras used in this paper.

%--------------------------------------------------
\subsection{$\mathfrak{sl}(2,\mathbb{R})\oplus\mathfrak{u}(1)$}
%--------------------------------------------------

In order for our notation to be as compact as possible we chose a basis such that the generators of $\mathfrak{sl}(2,\mathbb{R})\oplus\mathfrak{u}(1)$ denoted by  $\mathfrak{L}_n$ and $\mathfrak{S}$ with $n=0,\pm1$ satisfy
    \begin{subequations}
    \begin{align}
        [\mathfrak{L}_n,\mathfrak{L}_m] & = (n-m)\mathfrak{L}_{n+m},\\
        [\mathfrak{S},\mathfrak{L}_n] & = 0,\\
        [\mathfrak{S},\mathfrak{S}] & = 0.
    \end{align}
    \end{subequations}
Matrices representing these generators have been chosen to be real and in addition satisfy
    \begin{equation}
        \mathfrak{L}_n^\dagger=(-1)^n\mathfrak{L}_{-n}.
    \end{equation}
In terms of $3\times3$ matrices these generators read
    \begin{equation}
        \mathfrak{L}_1=\left(
			\begin{array}{ccc}
				0&0&0\\
				1&0&0\\
				0&0&0
			\end{array}\right),\quad
		\mathfrak{L}_0=\left(\begin{array}{ccc}
				\frac{1}{2}&0&0\\
				0&-\frac{1}{2}&0\\
				0&0&0
			\end{array}\right),\quad
		\mathfrak{S}=\left(\begin{array}{ccc}
				0&0&0\\
				0&0&0\\
				0&0&1
			\end{array}\right).
    \end{equation}
The corresponding bilinear form is proportional to the usual matrix trace and given by
	\begin{equation}
		\langle \mathfrak{L}_n\mathfrak{L}_m\rangle=\left(
			\begin{array}{c|ccc}
				  &\mathfrak{L}_1&\mathfrak{L}_0&\mathfrak{L}_{-1}\\
				\hline
				\mathfrak{L}_1&0&0&1\\
				\mathfrak{L}_0&0&-\frac{1}{2}&0\\
				\mathfrak{L}_{-1}&1&0&0
			\end{array}\right),\qquad \langle\mathfrak{S}\mathfrak{S}\rangle=1.
	\end{equation}
	
%--------------------------------------------------
\subsection{$\mathfrak{isl}(2,\mathbb{R})\inplus\mathfrak{u}(1)\inplus\mathfrak{u}(1)$}\label{sec:isl2u1u1}
%--------------------------------------------------

In order to construct matrix representations for $\mathfrak{isl}(2,\mathbb{R})\inplus\mathfrak{u}(1)\inplus\mathfrak{u}(1)$ one first can introduce the matrix $\gamma^\star_{(D)}$ as \cite{Gary:2014ppa,Riegler:2016hah}
	\begin{equation}\label{eq:GammaStarI}
	\gamma^\star_{(D)}=\left(
		\begin{array}{cc}
			\unity_{D\times D} & 0 \\
			0 & -\unity_{D\times D}\\
		\end{array}\right),
	\end{equation}
as well as a Grassmann parameter\footnote{The usefulness of such a parameter in the context of flat space holography has been first described in \cite{Krishnan:2013wta} and subsequently used in \cite{Gary:2014ppa,Riegler:2016hah} to construct explicit matrix representations of $\mathfrak{isl}(N,\mathbb{R})$.} $\epsilon$ that satisfies $\epsilon^2=0$. Using these two ingredients as well as the matrix representation for $\mathfrak{sl}(2,\mathbb{R})\oplus\mathfrak{u}(1)$ presented in the previous section one can write down a matrix representation where we have again
	\begin{equation}
		\Lt_n^\dagger=(-1)^n\Lt_{-n},\qquad \Mt_n^\dagger=(-1)^n\Mt_{-n}.
	\end{equation}
These generators are explicitly given by
	\begin{subequations}
	\begin{align}
		\Lt_1 & =\mathfrak{L}_1\otimes\unity_{2\times2}, &
		\Lt_0 & =\mathfrak{L}_0\otimes\unity_{2\times2}, &
		\Jt & = \mathfrak{S}\otimes\unity_{2\times2},\\
	    \Mt_1 & =\epsilon\,\mathfrak{L}_1\otimes\gamma^\star_{(1)}, &
		\Mt_0 & =\epsilon\,\mathfrak{L}_0\otimes\gamma^\star_{(1)}, &
		\Pt & = \epsilon\,\mathfrak{S}\otimes\gamma^\star_{(1)},
	\end{align}
	\end{subequations}
and satisfy \eqref{eq:isl2RBasis}. The invariant bilinear form displayed in \eqref{eq:ISLInvBilForm} can be obtained from this matrix representation using the hatted trace introduced in \cite{Gary:2014ppa,Riegler:2016hah}\footnote{Please note that in order to be consistent with the invariant bilinear forms used in this work our definition for the hatted trace differs by a factor of $\tfrac{1}{4}$ in comparison to the definition found in \cite{Gary:2014ppa,Riegler:2016hah}.} and that is defined by
	\begin{equation}\label{eq:cp64}
		\langle \mathcal{G}_{a} \mathcal{G}_{b} \rangle = \widehat{\textrm{Tr}}\big(\mathcal{G}_{a} \mathcal{G}_{b}\big) := \frac{\extd}{\extd \epsilon}\,{\textrm{Tr}}\big(\mathcal{G}_{a} \mathcal{G}_{b}\gamma^\ast_{(D)}\big)\big|_{\epsilon=0},
	\end{equation}
for some set of generators $\mathcal{G}_{a}$ whose matrix representation is given in terms of $2D\times2D$ matrices. Similarly the deformed bilinear form found in \eqref{eq:InvBilFormModEinstein} is given by 
    \begin{equation}\label{eq:AppDeformedInvBilForm}
    \mu \textrm{Tr}\big(\mathcal{G}_{a} \mathcal{G}_{b}\big),
    \end{equation}
where one has to take into account that $\epsilon^2=0$.

%--------------------------------------------------
\subsection{$\mathfrak{sl}(m|n)$ and $\mathfrak{sl}(2|1)$}
%--------------------------------------------------

In this subsection we collect some useful facts and formulas regarding matrix representations of super Lie algebras $\mathfrak{sl}(m|n;\mathbb{C})$ with a special focus on $\mathfrak{sl}(2|1)$. This will prove useful later on when defining a suitable matrix representation for the Lie algebra \eqref{eq:isl21}.\\
The superalgebra $\mathfrak{sl}(m|n;\mathbb{C})$ is the set of all complex valued $(m+n)\times(n+m)$ matrices $M$ that have the form
    \begin{equation}
        M=\left(
            \begin{array}{c|c}
                A & B \\
                \hline
                C & D
            \end{array}\right),
    \end{equation}
are equipped with the supercommutator
    \begin{equation}\label{eq:SuperCommutator}
        [M,M'\}=\left(
            \begin{array}{c|c}
                AA'-A'A+BC'+B'C & AB'-A'B+BD'-B'D \\
                \hline
                CA'-C'A+DC'-D'C & CB'+C'B+DD'-D'D
            \end{array}\right),
    \end{equation}
and satisfy the supertraceless condition
    \begin{equation}\label{eq:Supertraceless}
        \textrm{sTr}(M):=\textrm{Tr}(A)-\textrm{Tr}(D)=0.
    \end{equation}
Elements of this algebra with $B=C=0$ are called bosonic wheres elements with $A=D=0$ are called fermionic.\\
Using these definitions the superalgebra $\mathfrak{sl}(2|1)$ is the set of all $(2+1)\times(1+2)$ matrices that are supertraceless and satisfy the relations \eqref{eq:isl21}. A matrix representation in terms of $3\times3$ matrices is given by
    \begin{subequations}
    \begin{align}
        \mathfrak{L}_1 & =\left(
			\begin{array}{ccc}
				0&0&0\\
				1&0&0\\
				0&0&0
			\end{array}\right),&
		\mathfrak{L}_0 & =\left(\begin{array}{ccc}
				\frac{1}{2}&0&0\\
				0&-\frac{1}{2}&0\\
				0&0&0
			\end{array}\right),&
		\mathfrak{S} & =\left(\begin{array}{ccc}
				1&0&0\\
				0&1&0\\
				0&0&2
			\end{array}\right),\\
		\mathfrak{G}^+_\frac{1}{2} & =\left(
			\begin{array}{ccc}
				0&0&0\\
				0&0&0\\
				\sqrt{2}&0&0
			\end{array}\right),&
		\mathfrak{G}^-_\frac{1}{2} & =\left(\begin{array}{ccc}
				0&0&0\\
				0&0&\sqrt{2}\\
				0&0&0
			\end{array}\right),&
		 & 
    \end{align}
    \end{subequations}
where for the sake of compact presentation we have chosen a basis of real valued matrices such that
	\begin{equation}
		\mathfrak{L}_n^\dagger=(-1)^n\mathfrak{L}_{-n},\qquad \left(\mathfrak{G}^\pm_r\right)^\dagger=(-1)^{\frac{1}{2}\mp r}\mathfrak{G}^\mp_{-r}.
	\end{equation}
The bilinear form on this algebra is then given in terms of the supertrace i.e. ${\langle\ldots\rangle\equiv\textrm{sTr}[\ldots]}$ as
	\begin{equation}
		\langle \mathfrak{L}_n\mathfrak{L}_m\rangle=\left(
			\begin{array}{c|ccc}
				  &\mathfrak{L}_1&\mathfrak{L}_0&\mathfrak{L}_{-1}\\
				\hline
				\mathfrak{L}_1&0&0&1\\
				\mathfrak{L}_0&0&-\frac{1}{2}&0\\
				\mathfrak{L}_{-1}&1&0&0
			\end{array}\right),\quad
		\langle \mathfrak{G}^a_r\,\mathfrak{G}^b_s\rangle =\left(
			\begin{array}{c|cccc}
                &\mathfrak{G}^+_\frac{1}{2}&\mathfrak{G}^+_{-\frac{1}{2}}&\mathfrak{G}^-_\frac{1}{2}&\mathfrak{G}^-_{-\frac{1}{2}}\\
				\hline
				\mathfrak{G}^+_\frac{1}{2} & 0 & 0 & 0 & -1 \\
				\mathfrak{G}^+_{-\frac{1}{2}} & 0 & 0 & 1 & 0 \\
				\mathfrak{G}^-_\frac{1}{2} & 0 & -1 & 0 & 0 \\
				\mathfrak{G}^-_{-\frac{1}{2}} & 1 & 0 & 0 & 0
			\end{array}\right),
	\end{equation}
and $\langle\mathfrak{S}\mathfrak{S}\rangle=-2$.

%--------------------------------------------------
\subsection{$\mathfrak{isl}(m|n)$}\label{sec:isl21}
%--------------------------------------------------

After having recalled the definition of $\mathfrak{sl}(2|1)$ we will now define what we mean by the Lie superalgebra $\mathfrak{isl}(m|n)$ used in this paper. By this we mean the set of all complex valued supertraceless $2(m+n)\times2(n+m)$ matrices $\Lt$ and $\Mt$ that have the following form:
    \begin{equation}
        \Lt=\left(
            \begin{array}{c|c}
               \mathfrak{L} & 0 \\
                \hline
                0 & \mathfrak{L}
            \end{array}\right),\qquad
        \Mt=\epsilon\left(
            \begin{array}{c|c}
               \mathfrak{L} & 0 \\
                \hline
                0 & -\mathfrak{L}
            \end{array}\right),    
    \end{equation}
where $\mathfrak{L}\in\mathfrak{sl}(m|n)$ and $\epsilon$ is the nilpotent Grassmann parameter already introduced previously in Section~\ref{sec:isl2u1u1}. Or using the same notation as in Section~\ref{sec:isl2u1u1} it is the set of all matrices that can be constructed as a tensor product using $\mathfrak{L}\in\mathfrak{sl}(m|n)$, $\unity_{2\times2}$ as well as $\epsilon\gamma^\star_{(1)}$ in the following way:
    \begin{equation}
        \Lt:=\mathfrak{L}\otimes\unity_{2\times2},\qquad\Mt:=\epsilon\mathfrak{L}\otimes\gamma^\star_{(1)},
    \end{equation}
where $\Lt,\Mt\in\mathfrak{isl}(m|n)$.\\
One of the main advantages of this construction\footnote{It should be noted that this construction is closely related to what was called a ``despotic'' limit in \cite{Lodato:2016alv} and is one possible \.In\"on\"u--Wigner contraction of $\mathfrak{sl}(m|n)\oplus\mathfrak{sl}(m|n)$.} is that the supercommutator \eqref{eq:SuperCommutator} generalizes straightforwardly as
    \begin{equation}
        [\Lt,\Lt'\}=\left(
            \begin{array}{c|c}
                [\mathfrak{L},\mathfrak{L}'\} & 0 \\
                \hline
                0 & [\mathfrak{L},\mathfrak{L}'\}
            \end{array}\right),\quad
        [\Lt,\Mt'\}=\epsilon\left(
            \begin{array}{c|c}
                [\mathfrak{L},\mathfrak{L}'\} & 0 \\
                \hline
                0 & -[\mathfrak{L},\mathfrak{L}'\}
            \end{array}\right),\quad
        [\Mt,\Mt'\}=0.
    \end{equation}
In addition one can also straightforwardly define a ``super'' version of the hatted trace\footnote{This is not only possible for the hatted trace but all the traces defined in \cite{Gary:2014ppa,Riegler:2016hah}.} \eqref{eq:cp64} denoted by $\widehat{\textrm{sTr}}$ simply by replacing $\textrm{Tr}$ with $\textrm{sTr}$ in the definition \eqref{eq:cp64}. This super hatted trace can then be used to determine \eqref{eq:isl21InvBilinForm} using the matrix representation presented in this subsection. As in the $\mathfrak{isl}(2,\mathbb{R})$ case one can determine the deformed bilinear form \eqref{eq:InvBilFormModSUSY} again by replacing $\textrm{Tr}$ with $\textrm{sTr}$ in \eqref{eq:AppDeformedInvBilForm}.

%--------------------------------------------------
\section{Canonical Analysis}\label{sec:CanonicalAnalysis}
%--------------------------------------------------

%--------------------------------------------------
\subsection{A Gravity Toy Model}\label{sec:CAToy}
%--------------------------------------------------

For the sake of compactness we will in what follows only describe the canonical analysis of the boundary conditions for the unbarred gauge connections, since the procedure for the barred sector works exactly in the same way as the unbarred sector by simply replacing unbarred with barred quantities and exchanging $x^+\rightarrow x^-$ as well as taking into account the relative minus sign in front of the Chern-Simons action.\\
In order to determine the gauge transformations preserving the boundary conditions \eqref{eq:TOYBCs} we make an ansatz of the form
	\begin{subequations}\label{eq:ToyGaugeParameter}
	\begin{align}
		\epsilon(r,t,\varphi)&=b^{-1}\left[\sum\limits_{a=-1}^1\epsilon^a(x^+)\mathfrak{L}_a\right]b,\\
		\lambda(r,t,\varphi)&=\tilde{b}^{-1}\lambda(x^+)\mathfrak{S}\tilde{b}.
	\end{align}	
	\end{subequations}
In terms of this ansatz the gauge transformations the boundary condition preserving gauge transformations are given by
	\begin{equation}\label{eq:ToyBCPGTs}
		\epsilon^1=\epsilon_\mathcal{L},\quad \epsilon^0=-\epsilon_\mathcal{L}',\quad
		\epsilon^{-1}=-\frac{2\pi}{k}\left(\mathcal{L}-\frac{2\pi}{\kappa}\mathcal{K}^2\right)\epsilon_\mathcal{L}+\frac{\epsilon_\mathcal{L}''}{2},\quad
		\lambda=\epsilon_\mathcal{K}+\frac{4\pi}{\kappa}\epsilon_\mathcal{L},
	\end{equation}
where the functions $\epsilon_\mathcal{L}$ and $\epsilon_\mathcal{K}$ depend on $x^+$ and a prime denotes differentiation with respect to $\varphi$.\\
The fields $\mathcal{L}$ and $\mathcal{K}$ then transform under the gauge transformations \eqref{eq:ToyBCPGTs} as
	\begin{subequations}\label{eq:ToyGaugeTrafos}
	\begin{align}
		\delta_\epsilon\mathcal{L}&=\epsilon_\mathcal{L}\mathcal{L}'+2\mathcal{L}\epsilon_\mathcal{L}'-\frac{k}{4\pi}\epsilon_\mathcal{L}'''+\mathcal{K}\epsilon_\mathcal{K}',\\
		\delta_\epsilon\mathcal{K}&=\frac{\kappa}{4\pi}\epsilon_\mathcal{K}'+\epsilon_\mathcal{L}\mathcal{K}'+\mathcal{K}\epsilon_\mathcal{L}'.
	\end{align}
	\end{subequations}
The corresponding canonical boundary charges are obtained by functionally integrating \cite{Henneaux:1992,Blagojevic:2002aa}
    \begin{equation}
        \delta Q[\epsilon]+\delta Q[\lambda]=\frac{k}{2\pi}\int\extd\varphi\,\left\langle\epsilon \,\delta A_\varphi\right\rangle+\frac{\kappa_P}{4\pi}\int\extd\varphi\,\left\langle\epsilon \,\delta C_\varphi\right\rangle.
    \end{equation}
For the boundary conditions \eqref{eq:TOYBCs} one obtains 
    \begin{equation}\label{eq:ToyCanChargesVariation}
        \delta Q[\epsilon]+\delta Q[\lambda]=\int\extd\varphi\left(\epsilon_\mathcal{L}\delta\mathcal{L}+\epsilon_\mathcal{K}\delta\mathcal{K}\right),
    \end{equation}
which can be functionally integrated to yield 
    \begin{equation}\label{eq:ToyCanChargesEinstein}
        Q[\epsilon]+Q[\lambda]=\int\extd\varphi\left(\epsilon_\mathcal{L}\mathcal{L}+\epsilon_\mathcal{K}\mathcal{K}\right).
    \end{equation}
Combining the charges \eqref{eq:ToyCanChargesEinstein} with the infinitesimal transformations \eqref{eq:ToyGaugeTrafos} one finds the following non-vanishing Dirac brackets for the state dependent functions:
	\begin{subequations}
	\begin{align}
	\{\mathcal{L}(\varphi),\mathcal{L}(\bar{\varphi})\}_{\textrm{D.B}}&=2\mathcal{L}\delta'-\delta\mathcal{L}'-\frac{k}{4\pi}\delta''',\\
	\{\mathcal{L}(\varphi),\mathcal{K}(\bar{\varphi})\}_{\textrm{D.B}}&=\mathcal{K}\delta'-\delta\mathcal{K}',\\
	\{\mathcal{K}(\varphi),\mathcal{K}(\bar{\varphi})\}_{\textrm{D.B}}&=\frac{\kappa}{4\pi}\delta',
	\end{align}
	\end{subequations}
where all functions appearing on the r.h.s are functions of $\bar{\varphi}$ and prime denotes differentiation with respect to the corresponding argument. Moreover ${\delta\equiv\delta(\varphi-\bar{\varphi})}$ and $\delta'\equiv\partial_\varphi\delta(\varphi-\bar{\varphi})$. Expanding the fields and delta distribution in terms of Fourier modes as
	\begin{equation}\label{eq:ToyEinsteinFourierModes}
		\mathcal{L}=\frac{1}{2\pi}\sum\limits_{n\in\mathbb{Z}}\left(\mathfrak{L}_n-\frac{k}{4}\delta_{n,0}\right)e^{-in\varphi},\qquad
		\mathcal{K}=\frac{1}{2\pi}\sum\limits_{n\in\mathbb{Z}}\mathfrak{K}_ne^{-in\varphi},
	\end{equation}
and then replacing the Dirac brackets with commutators using $i\{\cdot,\cdot\}\rightarrow[\cdot,\cdot]$ one obtains the following non-vanishing commutation relations:
	\begin{subequations}
	\begin{align}
		[\mathfrak{L}_n,\mathfrak{L}_m]&=(n-m)\mathfrak{L}_{n+m}+\frac{c}{12}n(n^2-1)\delta_{n+m,0},\\
		[\mathfrak{L}_n,\mathfrak{K}_m]&=-m\mathfrak{K}_{n+m},\\
		[\mathfrak{K}_n,\mathfrak{K}_m]&=\frac{\kappa}{2}\,n\delta_{n+m,0},
	\end{align}
	\end{subequations}
where $c=6k$.

%--------------------------------------------------
\subsection{Einstein Gravity}\label{sec:CAEinstein}
%--------------------------------------------------

In this subsection we show how to compute the gauge transformations preserving \eqref{eq:NewFSBCs} as well as the associated canonical charges and the associated asymptotic symmetry algebra. In order to find these gauge transformations we make the ansatz
	\begin{subequations}\label{eq:GaugeParameter}
	\begin{align}
		\epsilon(r,u,\varphi)&=b^{-1}\left[\sum\limits_{a=-1}^1\epsilon^a(u,\varphi)\Lt_a+\sigma^a(u,\varphi)\Mt_a\right]b,\\
		\bar{\epsilon}(r,u,\varphi)&=\tilde{b}^{-1}\left[\bar{\epsilon}(u,\varphi)\Jt+\bar{\sigma}(u,\varphi)\Pt\right]\tilde{b}.
	\end{align}	
	\end{subequations}
In terms of this ansatz the gauge transformations (including proper and non-trivial ones) that preserve the boundary conditions \eqref{eq:NewFSBCs} are given by
	\begin{subequations}\label{eq:BCPGTs}
	\begin{align}
		\epsilon^1&=\epsilon_\mathcal{L},\quad \epsilon^0=-\epsilon_\mathcal{L}',\nonumber\\
		\epsilon^{-1}&=-\frac{\pi}{k}\left(\mathcal{M}-\frac{2\pi}{\kappa_P}\mathcal{P}^2\right)\epsilon_\mathcal{L}+\frac{\epsilon_\mathcal{L}''}{2},\\
		\sigma^1&=\sigma_\mathcal{M},\quad\sigma^0=-\sigma_\mathcal{M}',\nonumber\\
		\sigma^{-1}&=-\frac{\pi}{k}\left(\mathcal{N}-\frac{4\pi}{\kappa_P}\mathcal{J}\mathcal{P}\right)\epsilon_\mathcal{L}-\frac{\pi}{k}\left(\mathcal{M}-\frac{2\pi}{\kappa_P}\mathcal{P}^2\right)\sigma_\mathcal{M}+\frac{\sigma_\mathcal{M}''}{2},\\
		\bar{\epsilon}&=\epsilon_\mathcal{J}+\frac{4\pi}{\kappa_P}\epsilon_\mathcal{L}\mathcal{P},\quad\bar{\sigma}=\sigma_\mathcal{P}+\frac{4\pi}{\kappa_P}\left(\epsilon_\mathcal{L}\mathcal{J}+\sigma_\mathcal{M}\mathcal{P}\right),
	\end{align}
	\end{subequations}
where the functions $\epsilon_\mathcal{L}$, $\epsilon_\mathcal{J}$, $\sigma_\mathcal{M}$ and $\sigma_\mathcal{P}$ depend on $u$ and $\varphi$ and a prime denotes differentiation with respect to $\varphi$. In addition these functions have to satisfy
	\begin{equation}\label{eq:GaugeParametersTimeEvolution}
		\partial_u\epsilon_\mathcal{L}=\partial_u\epsilon_\mathcal{J}=0,\qquad\partial_u\sigma_\mathcal{M}=\partial_\varphi\epsilon_\mathcal{L},\qquad\partial_u\sigma_\mathcal{P}=\partial_\varphi\epsilon_\mathcal{J}\mathcal{K}.
	\end{equation}
That means that these gauge parameters can also be written as
    \begin{subequations}
    \begin{align}
    \epsilon_\mathcal{L} &=  \epsilon_\mathcal{L}(\varphi),& \sigma_\mathcal{M} &= \epsilon_\mathcal{M}(\varphi) +u\epsilon_\mathcal{L}' \\
    \epsilon_\mathcal{J} &= \epsilon_\mathcal{J}(\varphi),&\sigma_\mathcal{P} &= \epsilon_\mathcal{P}(\varphi) + u \epsilon_\mathcal{J}'.
    \end{align}
    \end{subequations} 
The fields $\mathcal{M}$, $\mathcal{L}$, $\mathcal{J}$ and $\mathcal{P}$ then transform under the gauge transformations \eqref{eq:BCPGTs} as
	\begin{subequations}\label{eq:GaugeTrafos}
	\begin{align}
		\delta_\epsilon\mathcal{M}&=\epsilon_\mathcal{L}\mathcal{M}'+2\mathcal{M}\epsilon_\mathcal{L}'+\mathcal{P}\epsilon_\mathcal{J}'-\frac{k}{2\pi}\epsilon_\mathcal{L}''',\\
		\delta_\epsilon\mathcal{L}&=\epsilon_\mathcal{M}\mathcal{M}'+2\mathcal{M}\epsilon_\mathcal{M}'-\frac{k}{2\pi}\epsilon_\mathcal{M}'''+\mathcal{P}\epsilon_\mathcal{P}'+\epsilon_\mathcal{L}\mathcal{L}'+2\mathcal{L}\epsilon_\mathcal{L}'+\mathcal{J}\epsilon_\mathcal{J}',\\
		\delta_\epsilon\mathcal{P}&=\epsilon_\mathcal{L}\mathcal{P}'+\mathcal{P}\epsilon_\mathcal{L}'+\frac{\kappa_P}{4\pi}\epsilon_\mathcal{J}',\\
		\delta_\epsilon\mathcal{J}&=\epsilon_\mathcal{M}\mathcal{P}'+\mathcal{P}\epsilon_\mathcal{M}'+\frac{\kappa_P}{4\pi}\epsilon_\mathcal{P}'+\epsilon_\mathcal{L}\mathcal{J}'+\mathcal{J}\epsilon_\mathcal{L}'.
	\end{align}
	\end{subequations}
The corresponding canonical boundary charges are obtained by functionally integrating \cite{Henneaux:1992,Blagojevic:2002aa}
    \begin{equation}
        \delta Q[\epsilon]+\delta Q[\bar{\epsilon}]=\frac{k}{2\pi}\int\extd\varphi\,\left\langle\epsilon \,\delta\mathcal{A}_\varphi\right\rangle+\frac{\kappa_P}{8\pi}\int\extd\varphi\,\left\langle\epsilon \,\delta\mathcal{C}_\varphi\right\rangle.
    \end{equation}
For the boundary conditions \eqref{eq:NewFSBCs} one obtains 
    \begin{equation}\label{eq:CanChargesVariation}
        \delta Q[\epsilon]+\delta Q[\bar{\epsilon}]=\int\extd\varphi\left(\epsilon_\mathcal{L}\delta\mathcal{L}+\epsilon_\mathcal{M}\delta\mathcal{M}+\epsilon_\mathcal{J}\delta\mathcal{J}+\epsilon_\mathcal{P}\delta\mathcal{P}\right),
    \end{equation}
which can be functionally integrated to yield 
    \begin{equation}\label{eq:CanChargesEinstein}
        Q[\epsilon]+Q[\bar{\epsilon}]=\int\extd\varphi\left(\epsilon_\mathcal{L}\mathcal{L}+\epsilon_\mathcal{M}\mathcal{M}+\epsilon_\mathcal{J}\mathcal{J}+\epsilon_\mathcal{P}\mathcal{P}\right).
    \end{equation}

%--------------------------------------------------
\subsection{$\mathcal{N}=4$ Supergravity}\label{sec:CASuper}
%--------------------------------------------------

In this subsection we determine the gauge transformations preserving \eqref{eq:NewSFSBCs}. The following analysis is conceptually very similar to the calculations performed in Section~\ref{sec:CAEinstein} with some minor differences that one has to be aware of.\\
In order to determine the gauge transformations that preserve the boundary conditions \eqref{eq:NewSFSBCs} we make a general ansatz for such a gauge parameter as
	\begin{equation}\label{eq:SuperGaugeParameter}
		\epsilon(r,u,\varphi)=b^{-1}\left[\sum\limits_{a=-1}^1\epsilon^a\Lt_a+\sigma^a\Mt_a+\sum\limits_{a=-\frac{1}{2}}^\frac{1}{2}{}^\pm\chi^a\Gt^\pm_a+{}^\pm\psi^a\Rt^\pm_a+\bar{\epsilon}\Jt+\bar{\sigma}\Pt\right]b,
	\end{equation}
where the functions $\epsilon^a$, $\sigma^a$, ${}^\pm\chi^a$, ${}^\pm\psi^a$, $\bar{\epsilon}$ and $\bar{\sigma}$ are arbitrary functions of $u$ and $\varphi$. The connection $\mathcal{A}$ transforms under such gauge transformations as
    \begin{equation}
        \delta_\epsilon\mathcal{A}=\extd\epsilon+[\mathcal{A},\epsilon\},
    \end{equation}
where $[,\}$ denotes the supercommutator that for $A$, $\bar{A}$ being bosonic and $B$, $\bar{B}$ being fermionic operatora acts as
    \begin{equation}
        [A,\bar{A}\}=[A,\bar{A}],\qquad [A,B\}=[A,B],\qquad [B,\bar{B}\}=\{B,\bar{B}\}.
    \end{equation}
In terms of this ansatz the gauge transformations that preserve the boundary conditions \eqref{eq:NewSFSBCs} are given by
	\begin{subequations}\label{eq:SuperBCPGTs}
	\begin{align}
		\epsilon^1&=\epsilon_\mathcal{L},\quad \epsilon^0=-\epsilon_\mathcal{L}',\nonumber\\
		\epsilon^{-1}&=-\frac{\pi}{k}\left(\mathcal{M}+\frac{\pi}{4k}\mathcal{P}^2\right)\epsilon_\mathcal{L}+\frac{\pi}{2k}\left(\mathcal{R}^+\epsilon_\mathcal{G}^++\mathcal{R}^-\epsilon_\mathcal{G}^-\right)+\frac{\epsilon_\mathcal{L}''}{2},\\
		\sigma^1&=\sigma_\mathcal{M},\quad\sigma^0=-\sigma_\mathcal{M}',\nonumber\\
		\sigma^{-1}&=-\frac{\pi}{k}\left(\mathcal{N}+\frac{\pi}{2k}\mathcal{J}\mathcal{P}\right)\epsilon_\mathcal{L}-\frac{\pi}{k}\left(\mathcal{M}+\frac{\pi}{4k}\mathcal{P}^2\right)\sigma_\mathcal{M}+\frac{\sigma_\mathcal{M}''}{2}\nonumber\\
		    &\quad+\frac{\pi}{2k}\left(\mathcal{G}^+\epsilon_\mathcal{G}^++\mathcal{G}^-\epsilon_\mathcal{G}^-+\mathcal{R}^+\epsilon_\mathcal{R}^++\mathcal{R}^-\epsilon_\mathcal{R}^-\right),\\
		{}^\pm\chi^\frac{1}{2} & = \epsilon_\mathcal{G}^\pm,\quad{}^\pm\chi^{-\frac{1}{2}} = -\frac{\pi}{2k}\left(\mathcal{R}^\mp\epsilon_\mathcal{L}\pm\mathcal{P}\epsilon_\mathcal{G}^\pm\right)-(\epsilon_\mathcal{G}^\pm)',\\
        {}^\pm\psi^\frac{1}{2} & = \epsilon_\mathcal{R}^\pm,\quad{}^\pm\psi^{-\frac{1}{2}} = -\frac{\pi}{2k}\left(\mathcal{R}^\mp\sigma_\mathcal{M}+\mathcal{G}^\mp\epsilon_\mathcal{L}\pm\mathcal{P}\epsilon_\mathcal{R}^\pm\pm\mathcal{J}\epsilon_\mathcal{G}^\pm\right)-(\epsilon_\mathcal{R}^\pm)',\\
		\bar{\epsilon}&=\epsilon_\mathcal{J}-\frac{\pi}{2k}\epsilon_\mathcal{L}\mathcal{P},\quad\bar{\sigma}=\sigma_\mathcal{P}-\frac{\pi}{2k}\left(\epsilon_\mathcal{L}\mathcal{J}+\sigma_\mathcal{M}\mathcal{P}\right),
	\end{align}
	\end{subequations}
where the functions $\epsilon_\mathcal{L}$, $\epsilon_\mathcal{J}$, $\epsilon_\mathcal{G}^\pm$, $\epsilon_\mathcal{R}^\pm$, $\sigma_\mathcal{M}$ and $\sigma_\mathcal{P}$ depend on $u$ and $\varphi$ and a prime denotes differentiation with respect to $\varphi$. In addition these functions have to satisfy
	\begin{equation}\label{eq:SuperGaugeParametersTimeEvolution}
		\partial_u\epsilon_\mathcal{L}=\partial_u\epsilon_\mathcal{J}=\partial_u\epsilon_\mathcal{G}^\pm=\partial_u\epsilon_\mathcal{R}^\pm=0,\quad\partial_u\sigma_\mathcal{M}=\partial_\varphi\epsilon_\mathcal{L},\quad\partial_u\epsilon_\mathcal{R}^\pm=\partial_\varphi\epsilon_\mathcal{G}^\pm,\quad\partial_u\sigma_\mathcal{P}=\partial_\varphi\epsilon_\mathcal{J}.
	\end{equation}
That means that these gauge parameters can also be written as
    \begin{subequations}
    \begin{align}
    \epsilon_\mathcal{L} &=  \epsilon_\mathcal{L}(\varphi),& \sigma_\mathcal{M} &= \epsilon_\mathcal{M}(\varphi) +u\epsilon_\mathcal{L}', \\
    \epsilon_\mathcal{R}^\pm &=  \epsilon_\mathcal{R}^\pm(\varphi),& \epsilon_\mathcal{G}^\pm &= \epsilon_\mathcal{G}^\pm(\varphi) +u(\epsilon_\mathcal{R}^\pm)', \\    
    \epsilon_\mathcal{J} &= \epsilon_\mathcal{J}(\varphi),&\sigma_\mathcal{P} &= \epsilon_\mathcal{P}(\varphi) + u \epsilon_\mathcal{J}'.
    \end{align}
    \end{subequations} 
The fields $\mathcal{M}$, $\mathcal{L}$, $\mathcal{J}$ and $\mathcal{P}$ then transform under the gauge transformations \eqref{eq:SuperBCPGTs} as
	\begin{subequations}\label{eq:SuperGaugeTrafos}
	\begin{align}
		\delta_\epsilon\mathcal{M}&=\epsilon_\mathcal{L}\mathcal{M}'+2\mathcal{M}\epsilon_\mathcal{L}'+\mathcal{P}\epsilon_\mathcal{J}'-\frac{k}{2\pi}\epsilon_\mathcal{L}'''\nonumber\\
		    &\quad-\frac{1}{2}\left(\epsilon_\mathcal{G}^+(\mathcal{R}^+)'+\epsilon_\mathcal{G}^-(\mathcal{R}^-)'+3\mathcal{R}^+(\epsilon_\mathcal{G}^+)'+3\mathcal{R}^-(\epsilon_\mathcal{G}^-)'\right),\\
		\delta_\epsilon\mathcal{L}&=\epsilon_\mathcal{M}\mathcal{M}'+2\mathcal{M}\epsilon_\mathcal{M}'-\frac{k}{2\pi}\epsilon_\mathcal{M}'''+\mathcal{P}\epsilon_\mathcal{P}'+\epsilon_\mathcal{L}\mathcal{L}'+2\mathcal{L}\epsilon_\mathcal{L}'+\mathcal{J}\epsilon_\mathcal{J}'\nonumber\\
		    &\quad-\frac{1}{2}\left(\epsilon_\mathcal{G}^+(\mathcal{G}^+)'+\epsilon_\mathcal{G}^-(\mathcal{G}^-)'+\epsilon_\mathcal{R}^+(\mathcal{R}^+)'+\epsilon_\mathcal{R}^-(\mathcal{R}^-)'\right.\nonumber\\
		    &\quad\left.+3\mathcal{G}^+(\epsilon_\mathcal{G}^+)'+3\mathcal{G}^-(\epsilon_\mathcal{G}^-)'+3\mathcal{R}^+(\epsilon_\mathcal{R}^+)'+3\mathcal{R}^-(\epsilon_\mathcal{R}^-)'\right),\\
		\delta_\epsilon\mathcal{R}^\pm & = -2\mathcal{M}\epsilon_\mathcal{G}^\pm\mathcal{R}^\pm\mathcal{J}\pm\epsilon_\mathcal{G}^\mp\mathcal{P}'\pm2\mathcal{P}(\epsilon_\mathcal{G}^\mp)'+\epsilon_\mathcal{L}(\mathcal{R}^\pm)'+\frac{3}{2}\mathcal{R}^\pm\epsilon_\mathcal{L}'+\frac{2k}{\pi}(\epsilon_\mathcal{G}^\mp)'',\\
		\delta_\epsilon\mathcal{G}^\pm & = -2\mathcal{N}\epsilon_\mathcal{G}^\pm-2\mathcal{M}\epsilon_\mathcal{R}^\pm\pm\mathcal{G}^\pm\mathcal{J}\pm\mathcal{R}^\pm\mathcal{P}\pm\epsilon_\mathcal{G}^\mp\mathcal{J}'\pm2\mathcal{J}(\epsilon_\mathcal{G}^\mp)'\pm\epsilon_\mathcal{R}^\mp\mathcal{P}'\pm2\mathcal{P}(\epsilon_\mathcal{R}^\mp)'\nonumber\\
		&\quad+\epsilon_\mathcal{L}(\mathcal{G}^\pm)'+\frac{3}{2}\mathcal{G}^\pm\epsilon_\mathcal{L}'+\epsilon_\mathcal{M}(\mathcal{R}^\pm)'+\frac{3}{2}\mathcal{R}^\pm\epsilon_\mathcal{M}'+\frac{2k}{\pi}(\epsilon_\mathcal{G}^\mp)'',\\ 
		\delta_\epsilon\mathcal{P}&=\epsilon_\mathcal{L}\mathcal{P}'+\mathcal{P}\epsilon_\mathcal{L}'+\epsilon_\mathcal{G}^+\mathcal{R}^+-\epsilon_\mathcal{G}^-\mathcal{R}^--\frac{k}{2\pi}\epsilon_\mathcal{J}',\\
		\delta_\epsilon\mathcal{J}&=\epsilon_\mathcal{M}\mathcal{P}'+\mathcal{P}\epsilon_\mathcal{M}'-\frac{k}{2\pi}\epsilon_\mathcal{P}'+\epsilon_\mathcal{L}\mathcal{J}'+\mathcal{J}\epsilon_\mathcal{L}'+\epsilon_\mathcal{G}^+\mathcal{G}^+-\epsilon_\mathcal{G}^-\mathcal{G}^-+\epsilon_\mathcal{R}^+\mathcal{R}^+-\epsilon_\mathcal{R}^-\mathcal{R}^-.
	\end{align}
	\end{subequations}
The corresponding canonical boundary charges are obtained by functionally integrating
    \begin{equation}
        \delta Q[\epsilon]=\frac{k}{2\pi}\int\extd\varphi\,\left\langle\epsilon \,\delta\mathcal{A}_\varphi\right\rangle.
    \end{equation}
For the boundary conditions \eqref{eq:NewSFSBCs} one obtains 
    \begin{equation}\label{eq:SuperCanChargesVariation}
        \delta Q[\epsilon]=\int\extd\varphi\left(\epsilon_\mathcal{L}\delta\mathcal{L}+\epsilon_\mathcal{M}\delta\mathcal{M}+\epsilon_\mathcal{G}^+\delta\mathcal{G}^++\epsilon_\mathcal{G}^-\delta\mathcal{G}^-+\epsilon_\mathcal{R}^+\delta\mathcal{R}^++\epsilon_\mathcal{R}^-\delta\mathcal{R}^-+\epsilon_\mathcal{J}\delta\mathcal{J}+\epsilon_\mathcal{P}\delta\mathcal{P}\right),
    \end{equation}
which can be functionally integrated to yield 
    \begin{equation}
        Q[\epsilon]=\int\extd\varphi\left(\epsilon_\mathcal{L}\mathcal{L}+\epsilon_\mathcal{M}\mathcal{M}+\epsilon_\mathcal{G}^+\mathcal{G}^++\epsilon_\mathcal{G}^-\mathcal{G}^-+\epsilon_\mathcal{R}^+\mathcal{R}^++\epsilon_\mathcal{R}^-\mathcal{R}^-+\epsilon_\mathcal{J}\mathcal{J}+\epsilon_\mathcal{P}\mathcal{P}\right).
    \end{equation}
    
%--------------------------------------------------
\section{Logarithmic Corrections via Saddle Point Approximation}
%--------------------------------------------------

Quantum gravity corrections in micro-canonical entropy of black holes generally behave logarithmically \cite{Banerjee:2010qc,Banerjee:2011jp,Sen:2011ba,Sen:2012cj,Sen:2012dw}. This follows from computing the quadratic fluctuations about the saddle used in getting the leading order behaviour from the density of states function.

In this direction, let us first denote the quantity in the exponential of the transform equation \eqref{eq:DOS} as:
\begin{equation}
f( \rho, \eta, \mu, \nu) = 2\left(\frac{\rho}{\eta^2}h_M^v-\frac{1}{\eta}h_L^v-\rho h_M-\eta h_L-\mu p-\nu j\right) - \frac{\nu^2}{2\eta}\kappa_J-\frac{2\eta\mu\nu-\nu^2\rho}{2\eta^2}\kappa_P .
\end{equation}
Expand it around the saddle defined by \eqref{eq:Saddle} up to second order:
\begin{equation} \label{eq:Fluct}
f(\vec{\lambda}) = f(\vec{\lambda_0}) + (\lambda^i -\lambda^i_0)(\lambda^j -\lambda^j_0) \partial_i \partial_j f(\vec{\lambda})\Big|_{\vec{\lambda} = \vec{\lambda_0}},
\end{equation}
where the quadruplet $\vec{\lambda}$ captures the modular and flow parameters in the order maintained above. In view of \eqref{eq:DOS}, $f(\vec{\lambda_0})$ serves as the leading term, whereas the contribution from quadratic fluctuation of \eqref{eq:Fluct} gives:
\begin{eqnarray}
d_{(2)} = \int \left(\Pi d\lambda_i \right) e^{\pi i (\lambda^i -\lambda^i_0)(\lambda^j -\lambda^j_0) \partial_i \partial_j f(\vec{\lambda})\Big|_{\vec{\lambda} = \vec{\lambda_0}}}.
\end{eqnarray}
Proper analytic continuation in the $\lambda$ space gives after us the relevant term for 
\begin{eqnarray}\label{eq:SaddleLog}
\log d_{(2)} &\sim & -\frac{1}{2}\log \left( \det \partial \partial f(\vec{\lambda})\Big|_{\vec{\lambda} = \vec{\lambda_0}}\right) \nonumber \\
&=& \log \left( \frac{h^v_M}{4 \kappa_P( \frac{p^2}{\kappa_P}-h_M)^2}\right) \nonumber \\
& \sim & -2 \log \left( \frac{12 p^2}{c_M} + h_M\right),
\end{eqnarray}
where $\partial \partial f(\vec{\lambda})$ stands for the $4 \times 4$ matrix that is formed by applying twice derivatives with respect to the four chemical potentials $\rho,\,\eta,\,\mu,\,\nu$. Note that this expression is spectral flow invariant, as expected.
\end{appendix}

%%%%%%%%%%%%%%%%%%%%%%%%%%%%%%%%%%%%%%%%%%%%%%%%%%%%%%%%%%%%%%%%%%%%%
%%%%%%%%%%%%%%%%%%%%%%%%%%%%% REFERENCES %%%%%%%%%%%%%%%%%%%%%%%%%%%%
%%%%%%%%%%%%%%%%%%%%%%%%%%%%%%%%%%%%%%%%%%%%%%%%%%%%%%%%%%%%%%%%%%%%%

\bibliographystyle{fullsort}
\bibliography{Bibliography}

\end{document}